\documentclass[12pt,article,5p,times,twocolumn,ieee]{elsarticle}

\usepackage[export]{adjustbox}
\usepackage{url}
\usepackage{amssymb}
\usepackage{float}
\usepackage{csquotes}
\usepackage{subfig}
\usepackage[percent]{overpic}
\usepackage{indentfirst}
\usepackage{fancyhdr}
\usepackage{mathtools} 
\usepackage{multirow}
\usepackage{graphicx}
\usepackage{caption,stackengine}
\usepackage{rotating} 
\usepackage{tikz-cd}
\usepackage{amsfonts}
\usepackage{amsthm}
\usepackage{graphicx,wrapfig,lipsum}
\usepackage{colortbl}
\usepackage{mathtools}
\usepackage{adjustbox}
\usepackage{pgfplots}
\pgfplotsset{width=10cm,compat=1.9}

\usepackage{fancyhdr}
\usepackage[utf8]{inputenc}
\usepackage[english]{babel}
\graphicspath{{Images/}}
\usepackage{subfig}
\usepackage{amsmath}

\usepackage{bm}
\usepackage{tabularx}
\usepackage{algorithm}
\usepackage{algorithmic}
\usepackage[normalem]{ulem}
\usepackage{enumitem}
\usepackage{amsthm,thmtools,xcolor} 
\usepackage{stfloats}
\usepackage{multirow}

\newcommand{\mathbbm}[1]{\text{\usefont{U}{bbm}{m}{n}#1}} 

\newcommand{\M}{\mathcal{M}}
\newcommand{\N}{\mathcal{N}}
\newcommand{\R}{\mathbbm{R}}
\newcommand{\F}{\mathcal{F}}
\newcommand{\myT}{\bar{T}}
\newcommand{\gtT}{T}
\newcommand{\ours}{\emph{ours}}

\newcommand{\dictionaryname}{\textsc{D}}
\newcommand{\ourname}{\textsc{PC}\dictionaryname}
\newcommand{\PCGAU}{\textsc{PC-Gau}}
\newcommand{\RECerr}{RECerr}

\renewcommand{\vec}{\mathbf}
\renewcommand{\epsilon}{\varepsilon}
\DeclareMathOperator{\geo}{GD}
\DeclareMathOperator*{\avg}{Avg}
\DeclareMathOperator*{\emb}{Emb}
\DeclareMathOperator*{\age}{AGE}

\DeclareMathOperator*{\re}{RE}
\DeclareMathOperator*{\fps}{FPS}
\DeclareMathOperator*{\pca}{PCA}

\DeclareMathOperator{\discr}{Dis}
\DeclareMathOperator{\egdc}{EGDC}
\DeclareMathOperator{\mgd}{MGD}
\DeclareMathOperator{\corr}{\rho}
\DeclareMathOperator*{\argmin}{argmin}

\newcommand\norm[1]{\left\lVert#1\right\rVert}
\newcommand{\tabred}[1]{\textcolor{red}{#1}}
\newcommand{\tabgreen}[1]{\textcolor{green}{#1}}
\newcommand{\tabblue}[1]{\textcolor{blue}{#1}}
\newcommand{\CGnew}[1]{\textcolor{black}{#1}}

\setlist[itemize,1]{label=--}
\setlist[itemize,2]{label=$\circ$}
\setlist[itemize,3]{label=$-$}

\journal{Computer \& Graphics}

\begin{document}

\lhead{}
\rhead{}
\chead{Accepted for Publication in Computer and Graphics, Elsevier https://doi.org/10.1016/j.cag.2023.04.010
}

\begin{frontmatter}



\title{Extracting a functional representation from a dictionary for non-rigid shape matching}

\author[]
{\parbox{\textwidth}{\centering Michele Colombo$^{1}$,
	  Giacomo Boracchi$^{1}$, Simone Melzi$^{2}$
      }
      \\{\parbox{\textwidth}{\centering $^1$Dipartimento di Elettronica e Informazione, Politecnico di Milano, Italy \\
$^2$Department of Informatics, Systems and Communication (DISCo), University of Milano-Bicocca, Italy 
}
}
}


\begin{abstract}
Shape matching is a fundamental problem in computer graphics with many applications. Functional maps translate the point-wise shape-matching problem into its functional counterpart and have inspired numerous solutions over the last decade. Nearly all the solutions based on functional maps rely on the eigenfunctions of the Laplace-Beltrami Operator (LB) to describe the functional spaces defined on the surfaces and then convert the functional correspondences into point-wise correspondences. However, this final step is often error-prone and inaccurate in tiny regions and protrusions, where the energy of LB does not uniformly cover the surface. We propose a new functional basis Principal Components of a Dictionary (\ourname{}) to address such intrinsic limitation. \ourname{} constructs an orthonormal basis from the Principal Component Analysis (PCA) of a dictionary of functions defined over the shape. These dictionaries can target specific properties of the final basis, such as achieving an even spreading of energy. Our experimental evaluation compares seven different dictionaries on established benchmarks, showing that \ourname{} is suited to target different shape-matching scenarios, resulting in more accurate point-wise maps than the LB basis when used in the same pipeline. This evidence provides a promising alternative for improving correspondence estimation, confirming the power and flexibility of functional maps.
\end{abstract}



\end{frontmatter}

\section{Introduction}
\label{sec:intro}%

\begin{figure*}[t]
    \centering
    {
    \subfloat[source]
    {\includegraphics[width=0.095\textwidth]{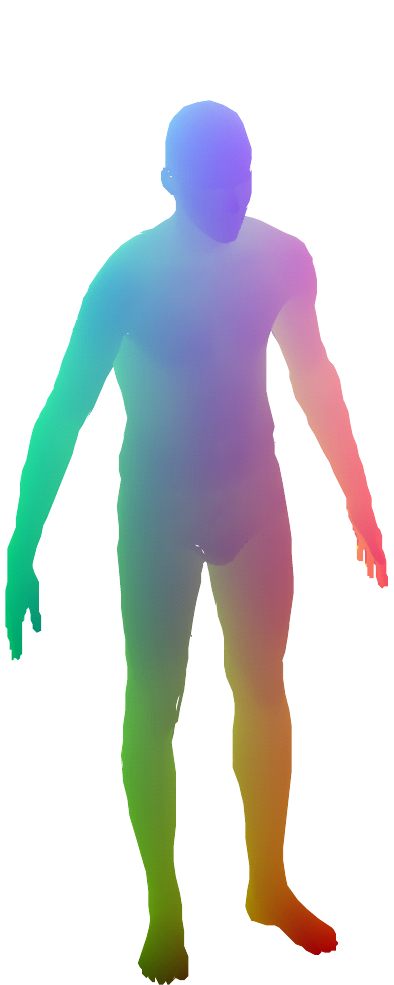}\label{sfig:matching:source}} \hfill
    \subfloat[point-wise maps]
    {\includegraphics[width=0.19\textwidth]{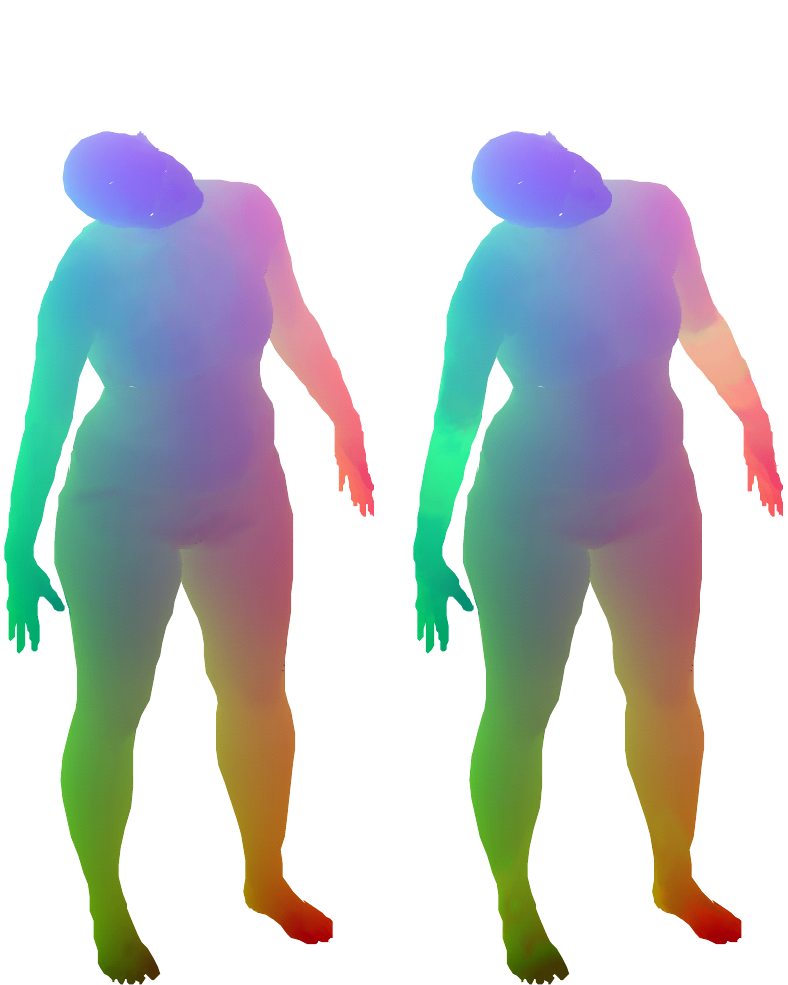}\label{sfig:matching:pMap}} \hfill
    \subfloat[geodesic error]{\includegraphics[width=0.24\textwidth]{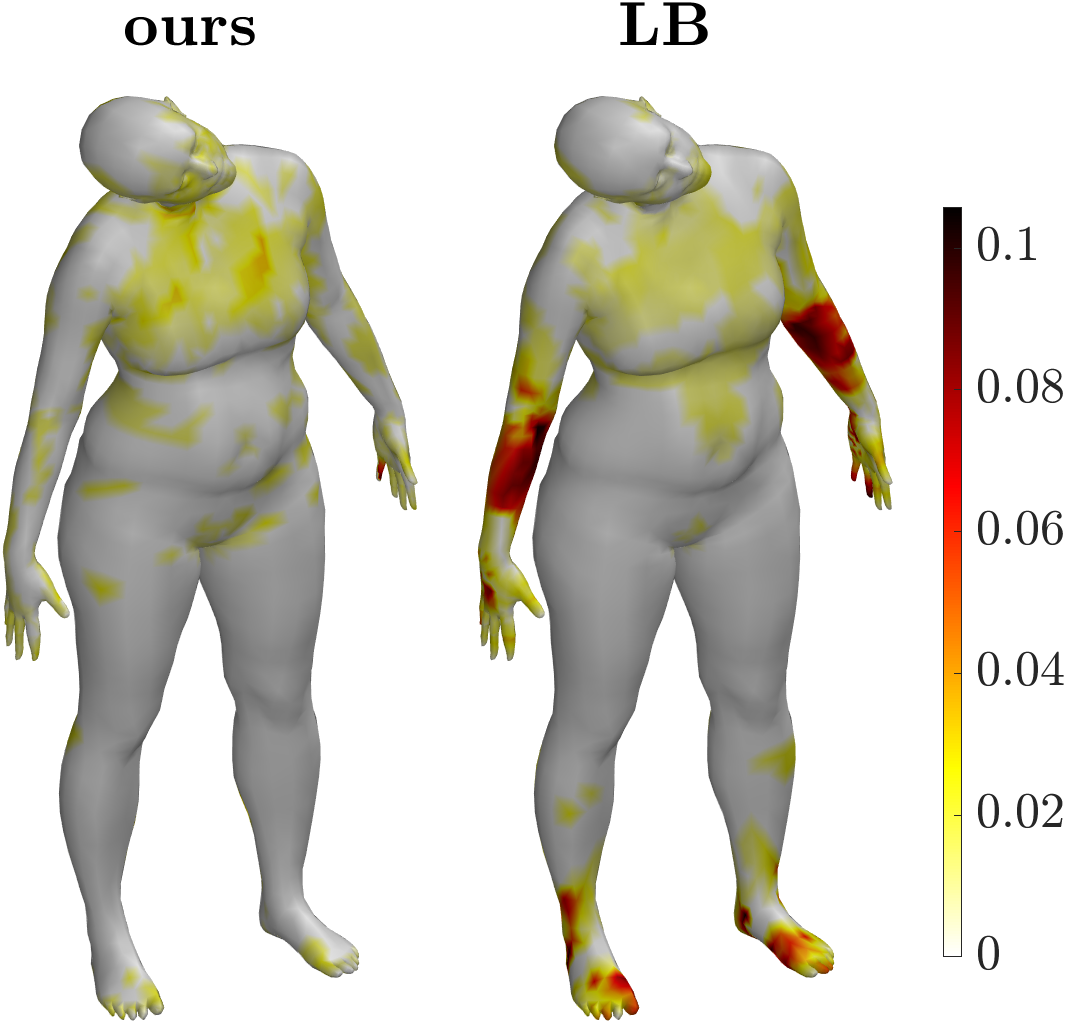}\label{sfig:matching:error}} 
    \hfill
    \subfloat[discrimination power]{\includegraphics[width=0.24\textwidth]{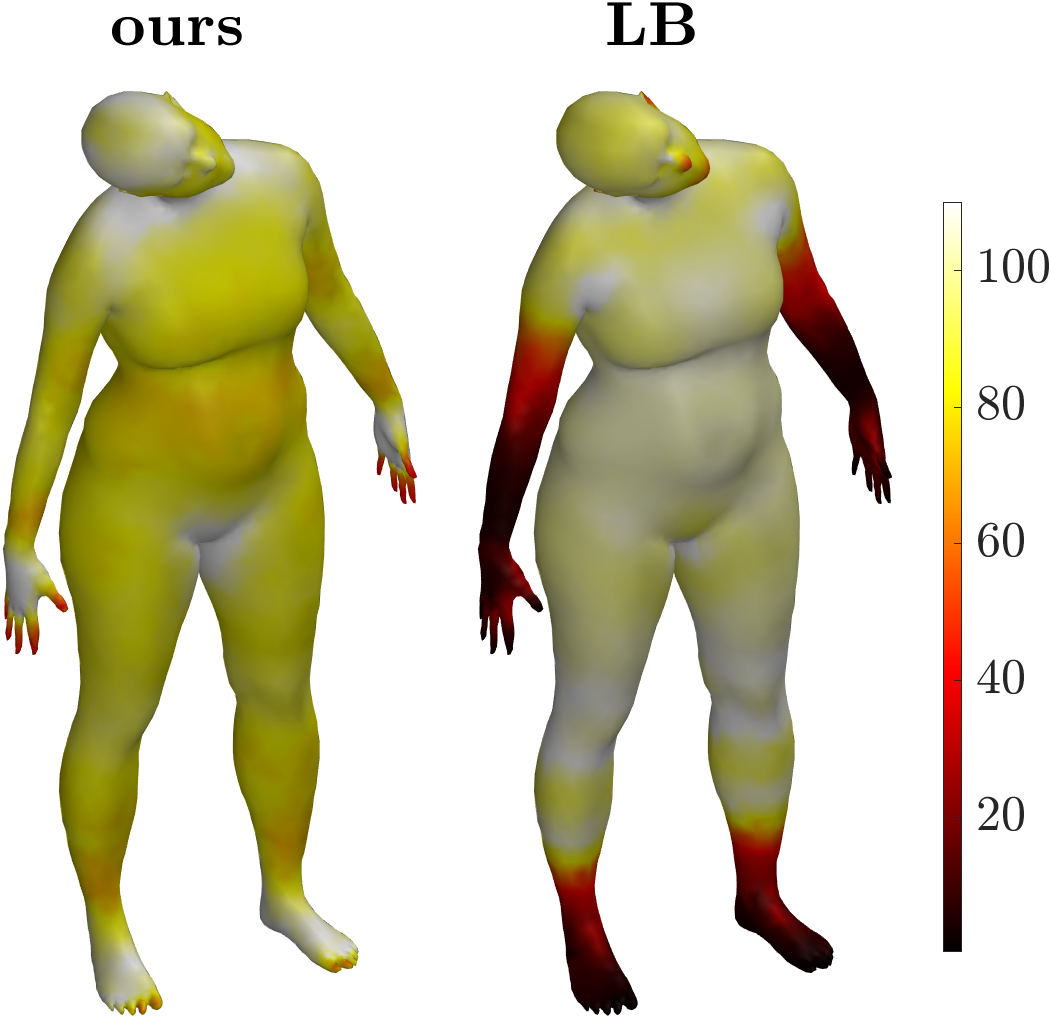}\label{sfig:matching:localization}} 
    }
    \caption{Example of shape matching between the source man \protect\subref{sfig:matching:source} and the target women \protect\subref{sfig:matching:pMap} comparing \ourname{} \CGnew{(ours with a dictionary of Gaussian)} and the standard basis Laplace Beltrami eigenfunctions (LB). \protect\subref{sfig:matching:pMap} the obtained point-wise maps between the woman and the man, encoded through color coding (correspondent points \CGnew{are depicted with} the same color). 
    \protect\subref{sfig:matching:error} the respective geodesic errors encoded by the colormap (white means \CGnew{perfect match} while dark colors are larger errors). \protect\subref{sfig:matching:localization} shows the spatial distribution of the energy of the basis. The energy of our basis is uniformly distributed on the mesh, which reflects in the error distribution.}
    \label{fig:matching_example}
\end{figure*}

Shape matching is a key problem in Computer Graphics and Geometry Processing. From an intuitive point of view, its goal is to estimate correspondences between points on a pair of 3D shapes, as shown in Figure~\ref{fig:matching_example}. Shape matching has a wide range of applications, including texture and deformation transfer~\cite{panozzo12fields, sumner2004deformation}, object retrieval~\cite{giorgi07SHREC07}, and statistical shape analysis~\cite{FAUST}. The problem is particularly challenging when a non-rigid deformation occurs between the shapes in the pair, such as a pose variation in human bodies.
In this case, the point-wise correspondence estimation is an NP-hard quadratic combinatorial problem in the number of points of the shapes.

Functional maps, the seminal work of Ovsjanikov \textit{et al.} \cite{ovsjanikov2012functional}, represented a breakthrough in the field. The key intuition behind functional maps is that it is easier to establish a correspondence among functions defined on the surfaces than directly finding a point-wise map between shapes. \CGnew{More specifically, functional maps builds upon a basis spanning a subspace of the functional space of each 3D shape. Then, functional correspondences can be described by a linear basis transformation, thus by a small matrix $C$, having dimensions equal to the number of basis functions in each shape. The matrix $C$} is estimated by solving an optimization problem exploiting linear constraints. Therefore, functional maps provides a compact representation for shape correspondence and, at the same time, a convenient tool to transfer functions from one shape to another. However, for many applications, it is necessary to recover a dense point-wise map~\cite{ezuz2017deblurring}, for instance, when we want to collect data variations in statistical shape analysis.

Recovering a point-wise map from a functional map is \CGnew{far from being straightforward}, and many works, starting from the seminal paper \cite{ovsjanikov2012functional}, have proposed different methods to accomplish this task. Recently, a few works tried to improve point-wise accuracy, focusing mainly on improving the estimation of matrix $C$~\cite{nogneng17, ren2018continuous, melzi2019zoomout}, or the algorithm applied to convert functional maps into point-wise maps~\cite{rodola15pointwise, ezuz2017deblurring, ren2018continuous, pai2021sikhorn}.

\CGnew{As a matter of fact, the choice of the basis turns to be key in this point-wise conversion as it } heavily affects the final result. The vast majority of methods, starting from~\cite{ovsjanikov2012functional}, have adopted the eigenfunctions of the Laplace-Beltrami operator (LB for brevity) to define the functional bases. These eigenfunctions are the equivalent of the harmonic basis~\cite{vallet08spectral} for non-Euclidean surfaces. \CGnew{Therefore, the} subset associated with the eigenvalues having the smallest absolute values is optimal for approximating smooth functions with limited variations~\cite{aflalo15optimality}. Nonetheless, LB presents a significant limitation for the point-wise conversion \CGnew{as LB functions} do not uniformly cover the mesh. In particular, the energy \CGnew{of LB functions} concentrates on the flat regions at the expense of narrower extremities \CGnew{and tiny protrusions}. In other words, LB's capability of discriminating between vertices and providing them with a meaningful representation is uneven across the surface. 

\CGnew{In the same spirit of previous works~\cite{kovnatsky2013coupled, neumann14compressed, LMH}, we propose an innovative framework to produce new bases for the functional spaces used in functional map pipelines. More specifically, we first generate a dictionary of \emph{informative} functions on the shape and then apply Principal Component Analysis (PCA) to obtain a compact set of orthonormal generators. The atoms of our basis are the Principal Components of a dictionary $\dictionaryname$, thus the name \ourname{}. The rationale behind \ourname{} is that by minimizing the reconstruction error on the atoms in $\dictionaryname$, the PCA produces a basis of a compact subspace that reflects the distribution of these functions in the initial dictionary. With \ourname{}, we can promote the resulting basis to be, for example, evenly distributed on the surface, which is difficult to achieve with LB. At the same time, \ourname{} is a compact set of orthonormal functions ready to use in the functional maps framework.}

\CGnew{We show that, for some of the dictionaries we propose, our framework can generate a rich functional representation for all the points of the surfaces. Specifically, we assess the quality of such representation by the following properties: (1) the capability of discriminating between different vertices and (2) the preservation of vertex locality. We provide a quantitative evaluation of such properties for the most effective dictionaries we have investigated and LB, showing a favorable comparison in this regard. At the same time, independently from the dictionary adopted,} our basis shares good properties with LB: it is orthonormal and invariant to isometries by construction and empirically exhibits a weak frequency order in their components, thus being a suitable and direct replacement for LB. These properties make our representation ready to be combined with other approaches proposed to improve the results achieved by the standard LB, tackling other aspects of the functional map framework and achieving significant benefits.

Through numerous experiments on established datasets, we show that the properties of our basis improve the quality of the final point-wise map. The similarity between our basis and LB enables us to directly compare them in the same setting. Thus, we test \ourname{} and LB when injected in different shape-matching pipelines~\cite{nogneng17,melzi2019zoomout} and also using functional maps computed from ground-truth correspondence. In these experiments, we see that our basis reaches significantly higher values of point-wise accuracy in all the analyzed settings. Our code is publicly available at \url{https://github.com/michele-colombo/PC-Gau_STAG2022}.

\CGnew{This work extends the idea proposed in \textit{``PC-GAU: PCA Basis of Scattered Gaussians for Shape Matching via Functional Maps''} \cite{stag2022}, where this procedure was applied only to a dictionary consisting of a collection of Gaussian functions scattered on the surface. Our major extensions with respect to \cite{stag2022} consists in generalizing the procedure and investigating different definitions of the initial dictionary (see Section \ref{sec:dictionaries}). Moreover, we assess the role and the actual dependence of \ourname{} from its more relevant parameters (Section \ref{sec:exp}) and we provide additional quantitative evaluation by expanding the test sets and scenarios of our experiments.}

\section{Related work} 
\label{sec:related_work}%
In this Section, we briefly overview the existing shape matching methods that fall, similarly to ours, into the functional map framework~\cite{ovsjanikov2012functional}. For other approaches to shape matching, refer to the survey~\cite{van2011survey,sahilliouglu2020recent}.

The functional map framework was proposed by Ovsjanikov \textit{et al.} in~\cite{ovsjanikov2012functional}. Its core idea is first to estimate the correspondence between functional spaces defined on the meshes and then extract a point-wise map from it. 
Given a basis for the functional space, the functional correspondence is compactly represented as a matrix $C$, which is estimated by imposing linear functional constraints. These constraints express correspondent landmarks, segments, and descriptors between the two shapes. The most used descriptors in this context are invariant to near-isometric deformations, like heat diffusion~\cite{sun2009concise} or quantum mechanical properties~\cite{aubry2011wave}.

Additional constraints can improve the estimation of the matrix $C$. Commutativity of the functional map with the LB operator, which forces the matrix $C$ to represent isometries, was firstly proposed in~\cite{ovsjanikov2012functional} and then refined in~\cite{ren2019structured}.
In~\cite{nogneng17}, the authors show that imposing the preservation of point-wise products is an additional, beneficial constraint. In shape with symmetries, the preferred maps do not mix up symmetries. This property has been promoted by introducing a new term in the optimization~\cite{ren2018continuous} or considering the complex counterpart of the functional map~\cite{donati2022complex}. An efficient method for converting a functional map into a point-wise map \CGnew{by exploiting an iterative closest point (ICP) as a form of refinement was} proposed in the original framework~\cite{ovsjanikov2012functional}. Subsequent works have proposed alternative solutions for extracting better point-wise maps from functional maps~\cite{rodola15pointwise}, by considering a probabilistic model~\cite{rodola2017regularized}, by introducing a smoothness prior on the point-wise map~\cite{ezuz2017deblurring}, or by devising a complex refinement scheme to promote continuity, coverage and bijectivity of the obtained map~\cite{ren2018continuous}. ZoomOut~\cite{melzi2019zoomout} is an iterative method for extending the size of an initial functional map estimated with few eigenvectors, \CGnew{while improving the quality of the estimated correspondence. Many recent algorithms build upon the ZoomOut procedure \cite{ruqi2020consistent,ren2020maptree,pai2021sikhorn,ren21discrete,panine22landmark}, which alternates conversions to point-wise maps and back to functional maps of increased size. } 
Despite pursuing the general goal of improving the quality of shape-matching via functional maps, all these methods differ radically from our approach since they focus either on the estimation of $C$ or on the process of extracting the point-wise map from $C$. They all make the implicit choice, ubiquitous since~\cite{ovsjanikov2012functional}, of using eigenfunctions of the Laplace-Beltrami operator~\cite{levy06laplace, vallet08spectral} as the functional basis, even though they are in principle agnostic to the basis adopted. This fact makes our proposal \emph{complementary} to most methods in the literature, 
as they can be used in combination with our basis. In perfect agreement with our intuition, 
\cite{nogneng18products} and~\cite{maggioli21orthogonalized} showed that extending the LB basis by adding point-wise products of atoms provides excellent benefits to the resulting point-wise maps. In principle, the same procedure can be applied to any basis, including ours.

A few other works followed an approach similar to ours, proposing a new bases for the functional space of a mesh. Some of them target specific tasks, like the transfer of tessellation structure~\cite{melzi20intrinsic} or step functions~\cite{melzi19step} from one mesh to another. The method in \cite{melzi20intrinsic} has also been used for shape matching, showing specific improvements in parts of the human body such as hands and feet. This solution, however, requires the meshes in the pair to be in similar poses, severely limiting its use in the general context. A different solution is pursued in \cite{neumann14compressed}, which proposes a basis whose atoms have local support, thus promoting the sparsity of $C$, but without improving the point-wise accuracy of the standard LB. Kovnatsky \textit{et al.}~\cite{kovnatsky2013coupled} built a coupled pair of bases by joint diagonalization to overcome the instability of LB in non-isometric pairs. Since the atoms of these coupled bases approximately diagonalize the Laplace-Beltrami operator, their energy will distribute similarly to the one of LB. 
Furthermore, a functional basis obtained with~\cite{kovnatsky2013coupled} also depends on the other mesh in the pair and the landmarks used. With a purpose more similar to ours, Melzi \textit{et al.}~\cite{LMH} proposed a basis that can extend LB and improve its expressive powers in specific mesh regions. However, their approach differs from ours: the atoms of their bases localize in predefined areas, which must be provided as input. This is a non-trivial information to be provided, as it requires the knowledge of at least coarse correspondences between the two shapes. On the contrary, our framework generally does not require any input other than the mesh itself. 

\CGnew{This work extends~\cite{stag2022}, which proposes to extract a new basis by PCA from a dictionary of Gaussian functions scattered uniformly on the shape. In this extended version, we improve the proposed procedure by considering a variety of dictionaries featuring different properties. Furthermore, we extend the analysis of the parameters of the method and evaluate the pipeline by considering additional test sets. Our novel contributions highlight the potential of extracting basis by PCA from a redundant dictionary in general. This paper aims to pave the way for further exploration of novel bases to improve the accuracy of the functional map framework.}

\section{Background}
\label{sec:background}%
In this section, we introduce some background notions and the notation we adopt in the following.

\subsection{Discrete surfaces}
In this paper, we refer to shapes as 2-dimensional surfaces embedded in $\mathbb{R}^3$.
In the continuous setting, we can represent these surfaces as a compact and connected smooth 2-dimensional Riemannian manifold $\M \subset \mathbb{R}^3 $. 
We refer to \cite{docarmo} for the definitions of differential geometry that are necessary to deal with this continuous representation of $\M$.
In the discrete setting we represent $\M$ as a triangular mesh $\M=(V_{\M}, E_{\M})$, where $V_{\M}$ is the set of $n$ vertices and $E_{\M}$ is the list of edges which means that $e_{xj} \in E_{\M} \iff $ exists an edge that connect $x$ and $j, \forall x, j \in V_{\M}$. \CGnew{We store the 3D coordinates of the vertices in $V_{\M}$ in a matrix $X_{\M} \in \R^{n\times 3}$. Each row of $X_{\M}$ corresponds to the position in the 3D space of a vertex of $\M$.}

\subsection{Shape matching}
\label{ssec:shape_matching}%
The input of shape matching is usually composed of two discrete meshes $\M$ and $\N$, having sets of vertices $V_{\M}$ and $V_{\N}$, respectively. Here we consider $|V_{\M}| = |V_{\N} | = n$ for simplicity of notation, but this is not a necessary assumption. We assume that an unknown correspondence $\gtT: V_{\N}\to V_{\M}$ between $\M$ and $\N$ exists without specific requirements on the function $\gtT$. For instance, $\M$ can be a non-rigid deformation of $\N$. We will refer to $\gtT$ as the ground truth point-wise map and we represent it either as a vector of vertex indices of size $n$ or as an $n \times n$ matrix $\Pi$ such that $\Pi_{ij} = 1$ if $\gtT(i)=j$ and $0$ otherwise, $\forall i \in V_{\N}$ and $\forall j \in V_{\M}$.

The goal of shape matching is, given $\M$ and $\N$, to estimate the unknown map $\gtT$. The estimated map $\myT$ has to be as close as possible to $\gtT$, which means that $\myT$ should assign to each vertex $y\in V_{\N}$ a vertex on $\M$ that is geodesically close, ideally coincident, to the one associated by $\gtT$. When the ground truth map $\gtT$ is available, we assess the mapping error of each vertex $y\in V_{\N}$ as:
\begin{equation}
\label{eq:geo_error}
    e(y) = \geo_{\M}(\myT(y), \gtT(y)),
\end{equation} 
where $\geo_{\M}$ is the geodesic distance on the surface $\M$.
Figure~\ref{sfig:matching:pMap} presents an example of two point-wise maps rendered through color correspondence, where we depicted correspondent points with the same color. 
Moreover, in Figure~\ref{sfig:matching:error}, we visualize their respective geodesic errors, encoded by the colormap where 0 error corresponds to white while dark colors denote larger ones.

Some shape matching pipelines require a set of input landmarks. A landmark is a couple of points $(y \in V_{\N}, x\in V_{\M})$ in known correspondence, namely $\gtT(y) = x$. In some of the experiments in Section~\ref{sec:exp}, we will refer to landmarks, which are not required to build the bases but only for the specific matching pipelines.

\subsection{Functional maps}
\label{ssec:fMaps}%
\begin{figure}[t]
    \centering
    \includegraphics[width=\linewidth]{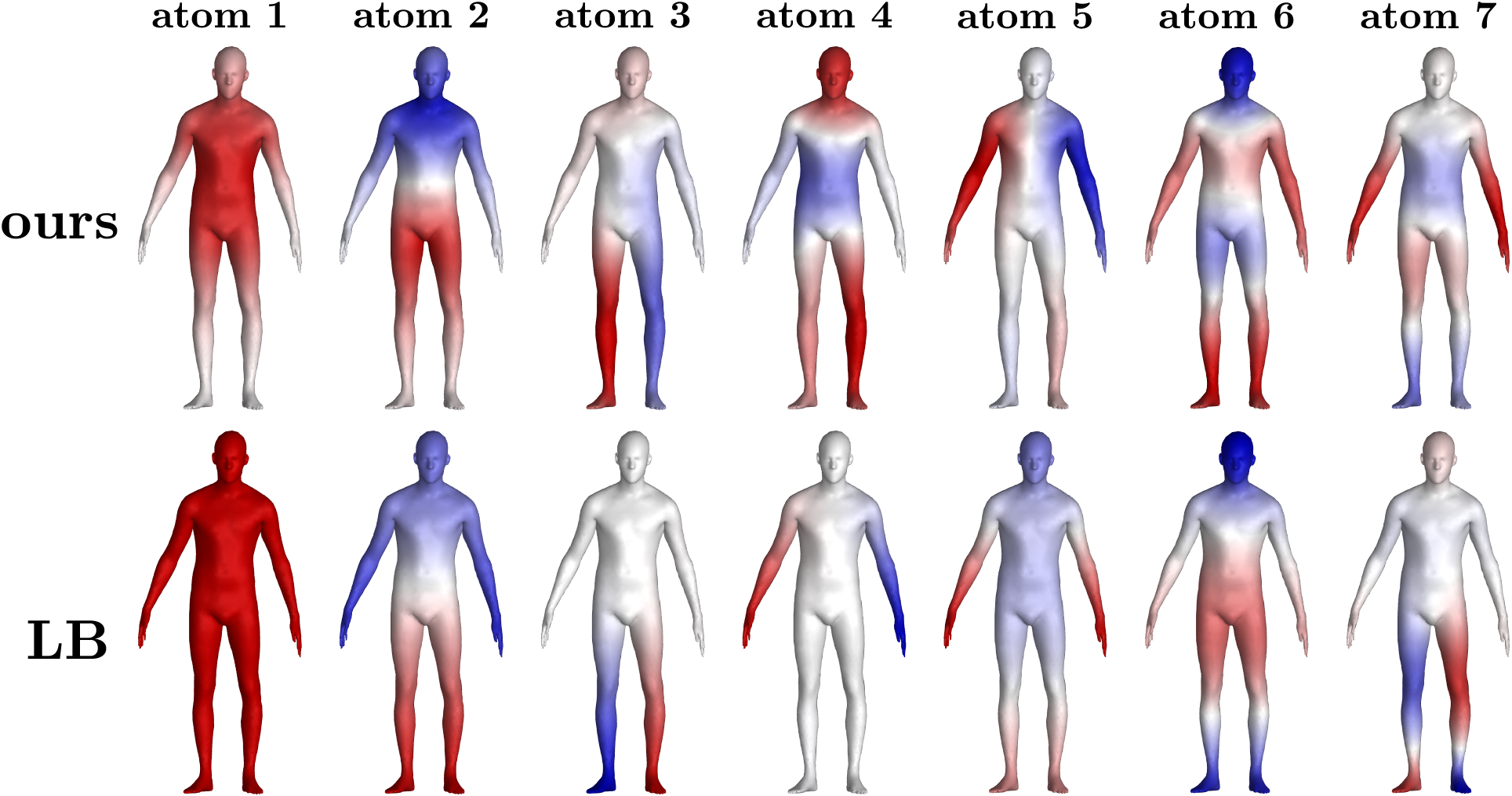}
    \caption{Atoms of our basis (top) compared to LB basis (bottom) atoms. Each element of the bases is a function defined on the mesh, represented here through color. Positive values are red, negative values are blue, and white corresponds to zero.}
    \label{fig:example_basis_atoms}
\end{figure}

\CGnew{In the discrete setting}, a real-valued function $f$ on $\M$ is given by a vector that associates to each vertex $x\in V_{\M}$ a value $f(x) \in \R$. We call $\F(\M,\R)$ the space of such functions. A basis for $\F(\M,\R)$ is a set of orthonormal functions belonging to $\F(\M,\R)$, an example of basis is shown in Figure~\ref{fig:example_basis_atoms}.
Given a pair of discrete surfaces $\M$ and $\N$ related by a ground truth point-wise map $\gtT:V_{\N}\to V_{\M}$, the functional map framework~\cite{ovsjanikov2012functional}, instead of estimating $\gtT$ directly, searches for a \CGnew{correspondence among functions defined on} $\M$ and $\N$. Then, once the \emph{functional correspondence} has been estimated, it extracts the corresponding point-wise map. 
 
Functional maps builds upon the observation that a point-wise map $\gtT$ induces a \emph{linear} operator $T_F:\F(\M,\R)\to\F(\N,\R)$ that maps functions defined on $\M$ to functions defined on $\N$ via the composition:
\begin{equation}
 T_F(f) = f \circ T \quad \quad  \forall f \in \F(\M,\R)
\end{equation}
Given a pair of bases $\Phi = \{\phi\}_i$ and $\Psi = \{\psi\}_j$ for $\F(\M,\R)$ and $\F(\N,\R)$, respectively, we can write
\begin{gather*}
    g = T_F(f) = T_F\Big(\sum_i a_i \phi_i\Big) = \sum_i a_i T_F\big(\phi_i\big) = \\
    = \sum_i a_i \sum_j c_{ji}\psi_j = \sum_{ji} a_i c_{ji} \psi_j = \sum_j b_j \psi_j
\end{gather*}
where $\vec{a} = [a_i]$ and $\vec{b} = [b_j]$ are the projections of $f$ and $g$ on $\Phi$ and $\Psi$ respectively. $c_{ji}$ is the projection of $T_F(\phi_i)$ on $\psi_j$ and depends only on $T_F$ and the two bases. Therefore $T_F$ is compactly represented by the matrix $C = [c_{ij}]$ and $\vec{b} = C\cdot \vec{a}$.

In practice, we consider only the first $k$ atoms of the bases, truncating the previous series after the first $k$ coefficients. As a matter of fact, $k$ is independent of the number of vertices $n$ of the meshes and usually $k\ll n$. Therefore, matching two shapes in the functional map framework \CGnew{boils down to the estimation of} a matrix $C$ of size $k\times k$. We can represent landmarks, corresponding segments, and descriptors as functions defined on $\M$ and $\N$ and find the functional map $C$ that best preserves the functional constraints (in the least square sense) introduced by these \CGnew{corresponding functions}. Note that, in principle, we can truncate the basis of $\M$ and $\N$ at a different number of atoms $k_{\M}$ and $k_{\N}$, thus producing a rectangular $C\in\R^{k_{\N} \times k_{\M}}$. However, for the sake of simplicity, we consider $k_{\M} = k_{\N} = k$ in this paper.

\paragraph*{Embedding} 
\label{par:embedding}%
Once we truncate a basis for $\F(\M, \R)$ to size $k$, we can store it in a matrix $\Phi_{\M}$, where each column is a basis atom represented as a vector of real values. $\Phi_{\M}$ has thus size $n\times k$, where $n$ is the number of vertices and $k$ is the number of basis atoms considered. We call \emph{spectral embedding} (or simply \emph{embedding}) of a vertex $x \in \M$, the vector of values assumed by all the basis functions in $x$: $\emb(x) = [\phi_i(x)]\in \R^k$. $\emb(x)$ also corresponds to the coefficients, in the basis $\Phi_{\M}$, of a Delta function centered in the vertex $x$. Thus $\Phi_{\M}^T$ contains the coefficients of all the Delta functions of $\M$ (one for each vertex) as column vectors.

\subsection{Conversion to point-wise map}
\label{ssec:p2p_conv}
\CGnew{Functional maps estimates a mapping between functional spaces built over $\M$ and $\N$, and this needs to be converted into a point-wise map $\myT:V_{\N}\to V_{\M}$,}. A simple solution, proposed in \cite{ovsjanikov2012functional}, consists in finding, for each column of $\Phi_{\N}^T$, the nearest neighbour in the columns of $C \Phi_{\M}^T$. This procedure corresponds to transfer Delta functions, and thus embeddings of vertices, from $\M$ to $\N$ through $C$ and putting similar points in the embedding space in correspondence.

\paragraph*{Summarizing} The complete functional-map pipeline for shape matching, as detailed in~\cite{ovsjanikov2012functional}, is: (1) find a truncated basis of size $k$ on each mesh, (2) find the $C\in\R^{k\times k}$ that best preserves some functional constraints, and (3) convert the functional map $C$ to a dense point-wise map $\myT$. In Figure~\ref{fig:pipeline}, we report an illustration of the entire procedure, focusing on step (1) since it is the one that our method addresses.

\subsection{Standard basis from LB Operator}
\label{ssec:LB_basis}%
The Laplace-Beltrami operator $\Delta_{\M}:\F(\M,\R)\to \F(\M,\R)$ associates to each function $f\in \F(\M,\R)$ another function that is the divergence of the gradient of $f$. For discrete meshes, this operator corresponds to a $n \times n $ matrix $\Delta_{\M}$ and is usually computed using the cotangent scheme~\cite{meyer2003discrete, pinkall93}. $\Delta_{\M}$ admits an eigendecomposition:
$\Delta_{\M} \phi_i = \lambda_i \phi_i$, where \CGnew{$\Phi = \{\phi_1, \phi_2, \ldots\}$} are its eigenfunctions with corresponding real eigenvalues $\Lambda = \{\lambda_1 \le \lambda_2 \le \dots{} \}$. $\Phi$ forms an orthonormal basis for $\F(\M,\R)$, namely LB. LB atoms are ordered in increasing frequencies which are encoded in the corresponding eigenvalue. We depict this order qualitatively in Figure~\ref{fig:example_basis_atoms} (bottom row) and quantitatively in Figure~\ref{fig:dirichlet_atoms}. Here, by frequency of a function $f\in\F(\M,\R)$, we intend the Dirichlet energy of $f$, which is a measure of its smoothness. LB is considered as the mesh equivalent of the harmonic basis~\cite{levy06laplace, vallet08spectral}. By selecting only the first $k$ atoms we obtain an optimal~\cite{aflalo15optimality} low-pass filter approximation of functions in $\F(\M,\R)$. From~\cite{ovsjanikov2012functional} on, LB has been the standard choice of the functional basis on meshes.

\subsection{Motivation}
\label{ssec:motivation}%
Despite its many strengths, LB presents a significant limitation: the basis energy is not evenly distributed on the mesh surface but is concentrated in massive areas, leaving narrower extremities less covered. To define the energy of a basis, we consider two properties in particular:
\begin{itemize}
    \item discrimination power between different vertices,
    \item locality preservation.
\end{itemize}
As we demonstrate through our experiments in Section~\ref{ssec:localization_energy}, the lack of these properties in some areas of the mesh produces bad assignments in the point-wise map, affecting its overall accuracy. Instead, the shape matching task requires a functional basis whose energy is evenly distributed on the mesh, i.e., the value of the previous two properties is similar across different mesh areas and better overall. 

\section{\CGnew{Proposed method}}
\label{sec:prop_sol}%
\CGnew{This section presents our core contribution PCD, namely Principal Component of a Dictionary: a new framework to build a basis for the functional space defined on a triangular mesh. The produced basis plays the role of the truncated basis in functional maps pipelines for shape matching. In other words, the output of PCD can be a replacement for LB to obtain more accurate point-wise maps.}

\CGnew{We first illustrate the proposed procedure to build the basis starting from a given dictionary of functions $D$. Then, in Section~\ref{sec:gaussian_dictionary}, we describe the procedure to construct the dictionary exploited in~\cite{stag2022} and analyze the properties that directly arise from its construction, paying particular attention to the \textit{desiderata} expressed in Section~\ref{ssec:motivation}. Finally, in Section~\ref{sec:dictionaries}, we propose six alternative dictionaries that can be compared to LB and to the solution from~\cite{stag2022}.}

\subsection{Building procedure}
\label{ssec:building_procedure}

\begin{algorithm}[t]
    \caption{\CGnew{The \ourname{} framework}}
    \label{alg:building_procedure}
    \begin{algorithmic}[1]
    \STATE \textbf{input:} $D$, $A_{\M}$, $k$, $normalize$
    \STATE $n = size(D,1) =$ number of vertices \\
    \STATE $q = size(D,2) =$ number of atoms \\
    \IF{$normalize$}
    \FOR{$i=\{1,\ldots, n\}$, $j = \{1, \ldots q \}$}
        \STATE $D_{ij} = D_{ij}/\sqrt{(D_{*j})^TA_{\M}D_{*j}}$ 
    \ENDFOR
    \ENDIF
    \STATE $P = \pca\left(D^T, \text{variableWeights = }A_{\M}\right)$
    \STATE $P_k = P(\texttt{:,1:}k)$
    \STATE \textbf{output: } $P_k$
    \end{algorithmic}
\end{algorithm} 

\begin{figure*}[!hbtp]
    \centering
    \includegraphics[width=\linewidth]{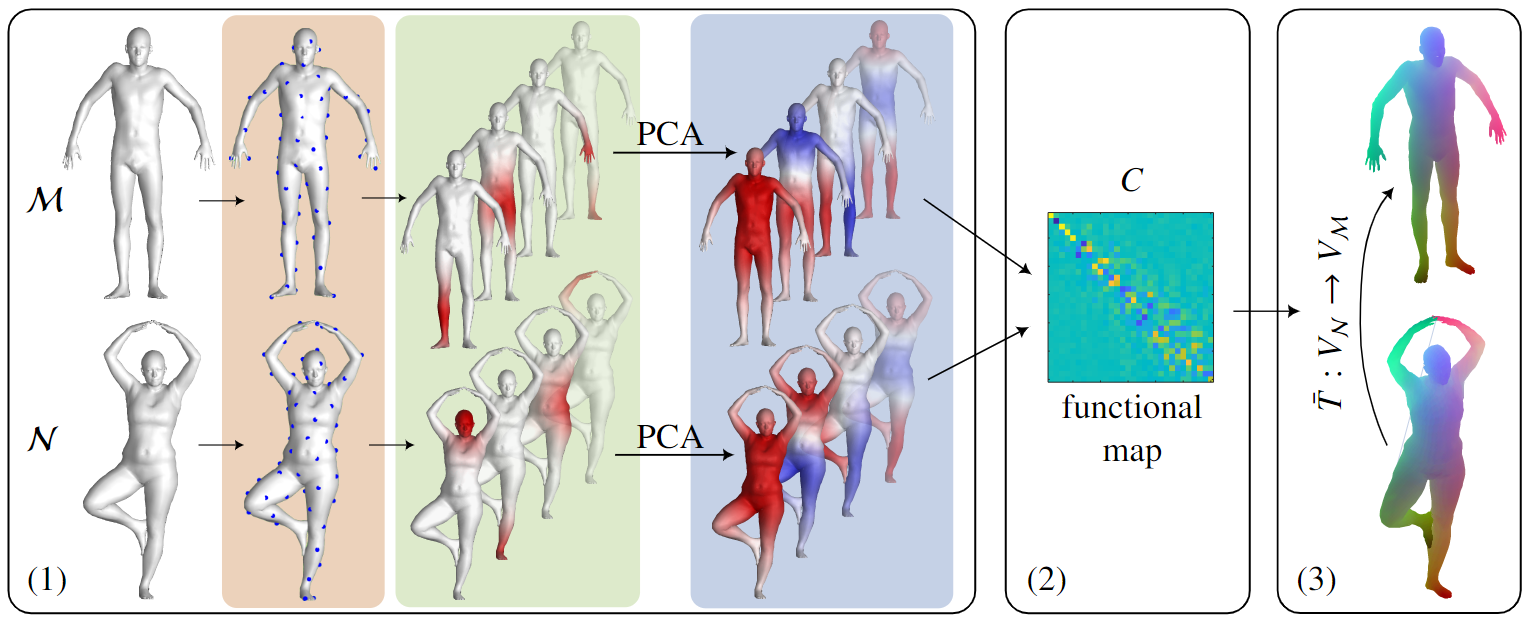}
    
    \caption{Complete shape matching pipeline with functional maps: (1) definition of a functional basis, (2) estimation of $C$ and (3) conversion to a point-wise map. Here, step (1) shows the building procedure \CGnew{proposed in~\cite{stag2022}}: selection of a subset of vertices (orange box), construction of the dictionary of Gaussian functions (green box) and dimensionality reduction through PCA (blue box).}
    \label{fig:pipeline}
\end{figure*}

\CGnew{Given a dictionary $D$ composed of $q\in\mathbb{N}$ functions, we perform dimensionality reduction through Principal Component Analysis (PCA), producing an orthogonal set of generators for the functional space spanned by the atoms in $D$.} Generators from PCA are ordered according to their capability to approximate the initial dictionary, therefore we select the only the first components. In the following, we present the pipeline to build our basis for any triangular mesh, \CGnew{which is summarized in Algorithm~\ref{alg:building_procedure}. In Figure~\ref{fig:pipeline} we report a scheme illustrating the entire pipeline, which includes the construction of the dictionary of Gaussians proposed in~\cite{stag2022}.}

\paragraph*{\CGnew{Input dictionary and first steps}}
Given a Mesh $\M$ with $n$ vertices, the input of our pipeline comprises two main ingredients and two parameters. The first ingredient is a dictionary $D \in \R^{n \times q}$, composed of $q$ functions defined on $\M$. The second one is $A_{\M}\in \R^{n\times n}$, the mass matrix of $\M$, which is a diagonal matrix whose entry $A_{\M}(i, i)$ corresponds to the area elements of the surface associated to the vertex $i$. The necessary parameters are an integer $k$ which sets the dimension of the generated basis, and $normalize$, a boolean variable which activates the normalization step.
$D$ is a fundamental, and almost the only choice in our pipeline. For this reason, we dedicate two sections to its definition, in which we show some possibilities and their properties. For now, let us consider a given, generic dictionary $D$, leaving deeper discussion on its choice to the following sections.

The algorithm starts by extracting the dimensionality  of the mesh, $n$ (line 2), and of the dictionary, $q$ (line3), from $D$.
In lines 4-8, if the variable $normalize$ is set to true, we normalize each column $D_{*j}$ of $D$ by division with its norm, computed as $\norm{D_{*j}}_{\M} = \sqrt{D_{*j}^T A_{\M} D_{*j}}$, where $A_{\M}$ defines the inner product on the functional space $\langle f, g \rangle_{\F(\M,\R)} = f^{\top}A_{\M}g$, $\forall f, g \in \F(\M,\R)$.

\paragraph*{Dimensionality reduction}
We then compute the PCA of \CGnew{$D^T$ (line 9),} meaning that \CGnew{each atom in the dictionary $D$} is considered a sample and each vertex of the mesh a variable. We do not centre the variables, but we weigh them for \CGnew{$A_{\M}(i, i)$}, the element of area associated with each vertex. The result of PCA is \CGnew{$P$,} a set of $q$ vectors of size $n$, called Principal Components (PCs). We can interpret these vectors as $q$ generators of the functional space $S$ spanned by \CGnew{the atoms in $D$}. Since $q<n$, $S$ is a proper subspace of $\F(\M,\R)$ and, \CGnew{if the atoms are linearly independent}, $S$ has dimension $q$.  

Finally, we select the first $k$ PCs to form our basis, and we store them in a matrix $P_k$ of size $n\times k$ \CGnew{(line 10)}. The columns of this matrix are a truncated basis of $\F(\M,\R)$ or, equivalently, a basis of a $k$-dimensional subspace $R\subset S \subset \F(\M,\R)$. In particular, PCA transform guarantees that the first $k$ PCs form the set of the $k$ orthonormal generators with the lowest approximation error on the initial samples~\cite{aflalo17regularizedPCA}:
\CGnew{
\begin{align}
\label{eq:pca_optimal}
    P_k = & \argmin_{P_k\in\R^{n\times k}} \left\{ \sum_{i=1}^q \norm{D_{*i} -  P_k P_k^T A_{\M}D_{*i}}_2^2 \right\} \\
    & \text{ s.t.} \quad  P_k^T A_{\M} P_k = I_k \nonumber
\end{align}
}
\CGnew{where $D_{*i}$ is the $i$-th column of $D$, namely the $i$-th atoms of the dictionary, and $I_k$ is the identity matrix of size $k$}. Orthonormality and projection in~\eqref{eq:pca_optimal} are expressed with respect to the inner product defined on the mesh \CGnew{$\langle \cdot , \cdot \rangle_{\F(\M,\R)}$}. \CGnew{We remark that the truncated basis $P_k$ satisfies Equation~\eqref{eq:pca_optimal}, thus provides a good approximation of $D$ by the spanned subspace $S$.}

\section{The dictionary of Gaussians}
\label{sec:gaussian_dictionary}

\CGnew{In~\cite{stag2022}, we consider a single choice of dictionary $D$, which we construct by taking inspiration from the signal processing literature~\cite{dapoian14gaussian}. More specifically, we consider a dictionary $D$ composed of a collection of Gaussian functions centred in a subset of vertices that uniformly cover the surface of the input mesh. Despite its simplicity, this basis enjoys many desirable properties for devising accurate shape correspondences via functional maps. Indeed, we show through a broad set of quantitative evaluations that the obtained basis is evenly distributed on the mesh surface and guarantees excellent discrimination power and locality preservation. In this section, we describe the procedure to define the dictionary of Gaussians and then analyze the properties that give rise to its superior accuracy in shape matching.
}

\subsection{\CGnew{Design of the dictionary of Gaussians}}
\CGnew{In the following, we present the steps required to construct the dictionary of Gaussians for a mesh $\M=(V_{\M}, E_{\M})$ with $n$ vertices. In Algorithm~\ref{alg:gaussians_dictionary}, we report in a compact format the proposed procedure which we represent visually on the left of Figure~\ref{fig:pipeline} (1).}

\begin{algorithm}[t]
    \caption{\CGnew{Construction of the dictionary of Gaussians}}
    \label{alg:gaussians_dictionary}
    \begin{algorithmic}[1]
    \STATE \textbf{input:} $\M$, $V_{\M}$, $\sigma$, $q$
    \STATE $Q = \fps(\M, q)$ \\
    \FOR{$i\in V_{\M}$, $j\in Q$} 
        \STATE $G_{ij} = \geo_{\M}(i,j)$
        \STATE $D_{ij} = \exp{\left(-G_{ij}^2/\sigma\right)}$
    \ENDFOR
    \STATE \textbf{output: } $D$
    \end{algorithmic}
\end{algorithm} 

\paragraph*{Subset of vertices} 
We start by selecting a subset \CGnew{$Q\subseteq V_{\M}$} of $q$ vertices on the mesh with Farthest Point Sampling~\cite{moenning03FPS} (line 2), using Euclidean distance for efficiency reasons. Thus is not the only viable critera, as discussed in Section~\ref{ssec:exp_gt}. The orange box in Figure~\ref{fig:pipeline} shows that the sampled points evenly cover the mesh surface. \CGnew{The idea is to obtain an even distribution of basis energy by controlling the uniformity of scattering of the Gaussians in the dictionary, since uniform sampling is rather easy to enforce.}

\paragraph*{Gaussian computation}
We compute $q$ Gaussian functions, each one centered in a vertex of $Q$. To do so, for each vertex $j\in Q$ (line 3) we compute the geodesic distance $G_{ij}=\geo_{\M}(i,j)$ to any other vertex $i\in V_{\M}$ (line 4) and then define the value of the Gaussian $D_{ij}$ (line 5) as:
\begin{align}
\label{eq:gaussian}
\CGnew{D_{ij}} &\CGnew{= \exp\left(\frac{-G_{ij}^2}{\sigma}\right)    .}
\end{align}
We approximate the geodesic distance as the length of the shortest path on the edges of the mesh~\cite{Mitchell87}. 
The parameter $\sigma$ is arbitrarily chosen and sets the amplitude of the Gaussians. We store the $\CGnew{q}$ Gaussian functions as columns of the matrix $\CGnew{D = [D_{*1}, D_{*2}, \ldots , D_{*q} ]}$, of size $n\times q$. 
In the green box in Figure~\ref{fig:pipeline}, we report some examples of the resulting Gaussian functions.
In all the experiments we use $q=1000$, $\sigma = 0.05$ and we apply normalization. 
\CGnew{In Section~\ref{ssec:param}, we analyze the impact of these two parameters on the accuracy of the point-to-point correspondence computed by our complete pipeline. These experiments were not included in~\cite{stag2022} and validate our choices. Once we have the dictionary $D$, we can apply the steps described in Section~\ref{sec:prop_sol} to obtain a basis of dimension $k$ for the functional space $\F(\M,\R)$. Coherently with \cite{stag2022}, we refer to this basis as \PCGAU{} or \ours{}.} 

\subsection{Properties of \PCGAU{}} The basis obtained in such a way shares many of the good properties of LB, which makes \PCGAU{} a suitable replacement for LB in existing functional map pipelines.

\begin{figure}[ht]
    \centering
    \includegraphics[width=\linewidth]{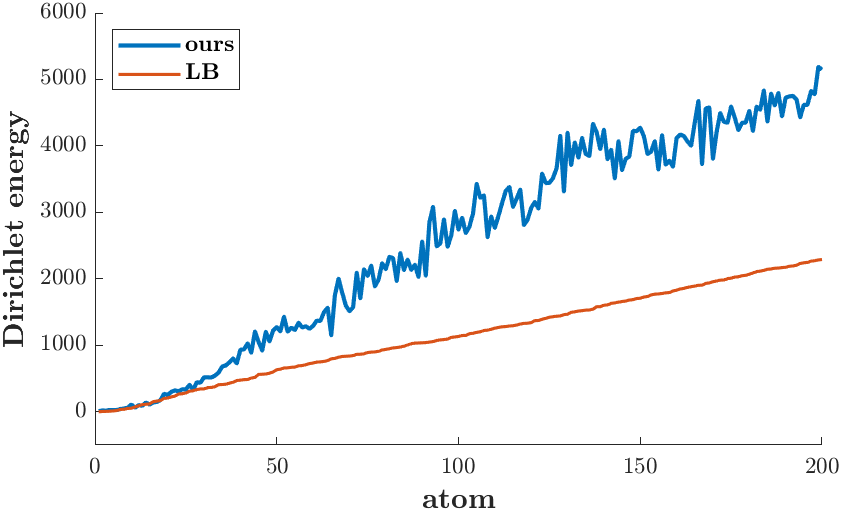}
    \caption{Comparison between the Dirichlet energy (frequency) of atoms of our basis and LB, computed for an example mesh from FAUST. Atoms of \PCGAU{} are approximately ordered by increasing frequency.}
    \label{fig:dirichlet_atoms}
\end{figure}

\subsubsection{Frequency ordering} Figure~\ref{fig:dirichlet_atoms} compares the Dirichlet energy of the atoms of \PCGAU{} and LB. The Dirichlet energy measures the smoothness of a function and can be interpreted as its frequency. We observe that, although not perfectly, atoms of \ours{} are approximately ordered in frequency. Even if not explicitly imposed, this order arises naturally in our basis as we empirically observe. We can also evaluate this qualitatively in the example atoms shown in Figure~\ref{fig:example_basis_atoms}. Frequency ordering means that, similarly to LB, we operate a low-pass filter approximation when we project a function on our truncated basis. We consider sorting our atoms with respect to their frequency as an interesting research direction. Still, we believe that their natural order is sufficient for our goal, and we leave its explicit constraint as future work.

\subsubsection{Orthonormality}
\label{ssec:orthonormality}
\PCGAU{}, like LB, is an orthonormal basis. We say that a basis $\Phi$ is orthonormal according to the inner product of \CGnew{the functional space $\F(\M,\R)$} if $\Phi^T A_{\M} \Phi = I_k$, where $A_{\M}$ is the mass matrix and $I_k$ is the identity matrix of size $k$. We can project a function $f$ on an orthonormal basis $\Phi_{\M}$ simply by matrix multiplication: $\vec{a} = \Phi^T_{\M} A_{\M} f$. Similarly, we can recover the function from its projection through $f = \Phi \vec{a}$. This is useful, in particular, when converting a given point-wise map $\Pi: \N \to \M$ (represented here as a matrix) to a functional map $C$, which boils down to: 
\begin{equation}
\label{eq:CfromGT}
    C = \Phi_{\N}^T A_{\M} \Pi \Phi_{\M} .
\end{equation}

\subsubsection{Isometry-invariance}
\begin{figure*}[!hbtp]
    \centering
    \includegraphics[width = 0.8\linewidth]{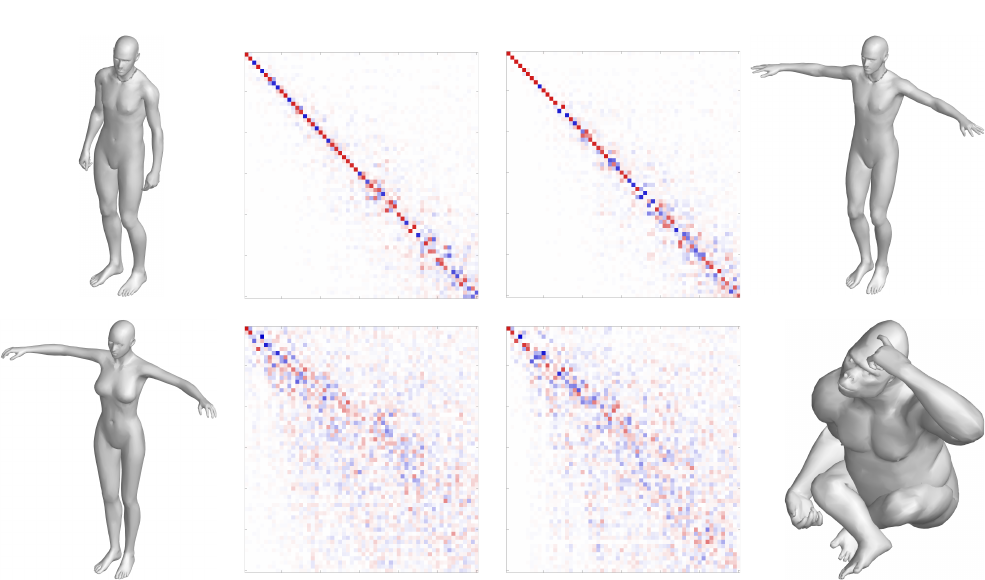}
    \caption{\CGnew{Example of functional maps in case of a near-isometric pair (top) and a non-isometric pair (bottom), computed using the ground-truth correspondence. By comparing the structure of the matrix for \PCGAU{} (left) and LB (right), we can see that their behaviour with respect to isometry is similar.}}
    \label{fig:isometry_invariance}
\end{figure*}

Two meshes $\M$ and $\N$ are isometric if the underlying correspondence $\gtT:V_{\N}\to V_{\M}$ is an isometry, which means that it preserves the geodesic distance of any pair of vertices on $\N$:
\begin{equation}
    \geo_{\M}(\gtT(x),\gtT(y)) = \geo_{\N}(x,y) \quad \forall x,y \in V_{\N}.
\end{equation}
Since we construct our basis purely on geodesic distances, under the assumption of equal distribution of the sample points, then our basis will be isometry-invariant. As for the LB, \CGnew{when shapes are isometric}, the functional map is a diagonal matrix with $+1$ or $-1$ on the diagonal. Most often, the energy of $C$ is funnel-shaped, approaching diagonal as the relation between $\M$ and $\N$ gets closer to isometry. 
\CGnew{In Figure~\ref{fig:isometry_invariance}, we show a qualitative comparison of the structure of the functional map computed with LB (left) and with \PCGAU{} (right) for a near-isometric (top row: two poses of the same human subject) and a non-isometric pair (bottom row: a woman and a gorilla in different poses). We remark that both bases in the two settings generate a matrix $C$ with a similar shape: close to diagonal for the near-isometric pair and a funnel shape for the non-isometric one. Their similarity further support our claim that the \PCGAU{} exhibits an invariance to isometry similar to the one of LB.}

\subsection{Spatial distribution of basis energy}
\label{ssec:localization_energy}%
We now consider the spatial distribution of the energy of the basis on the mesh surface. Intuitively, by basis energy in a vertex $x$, we mean the expressive power and the quality of the embedding space in a neighborhood of $x$. More precisely, as we stated in Section~\ref{ssec:motivation}, we introduce two quantities to assess the quality of the embedding space: (1) the discrimination power and (2) the preservation of locality.

\subsubsection{Discrimination power}
\label{ssec:discrimination_localization}%
\begin{figure}[ht]
    \centering
   \includegraphics[width=0.95\linewidth]{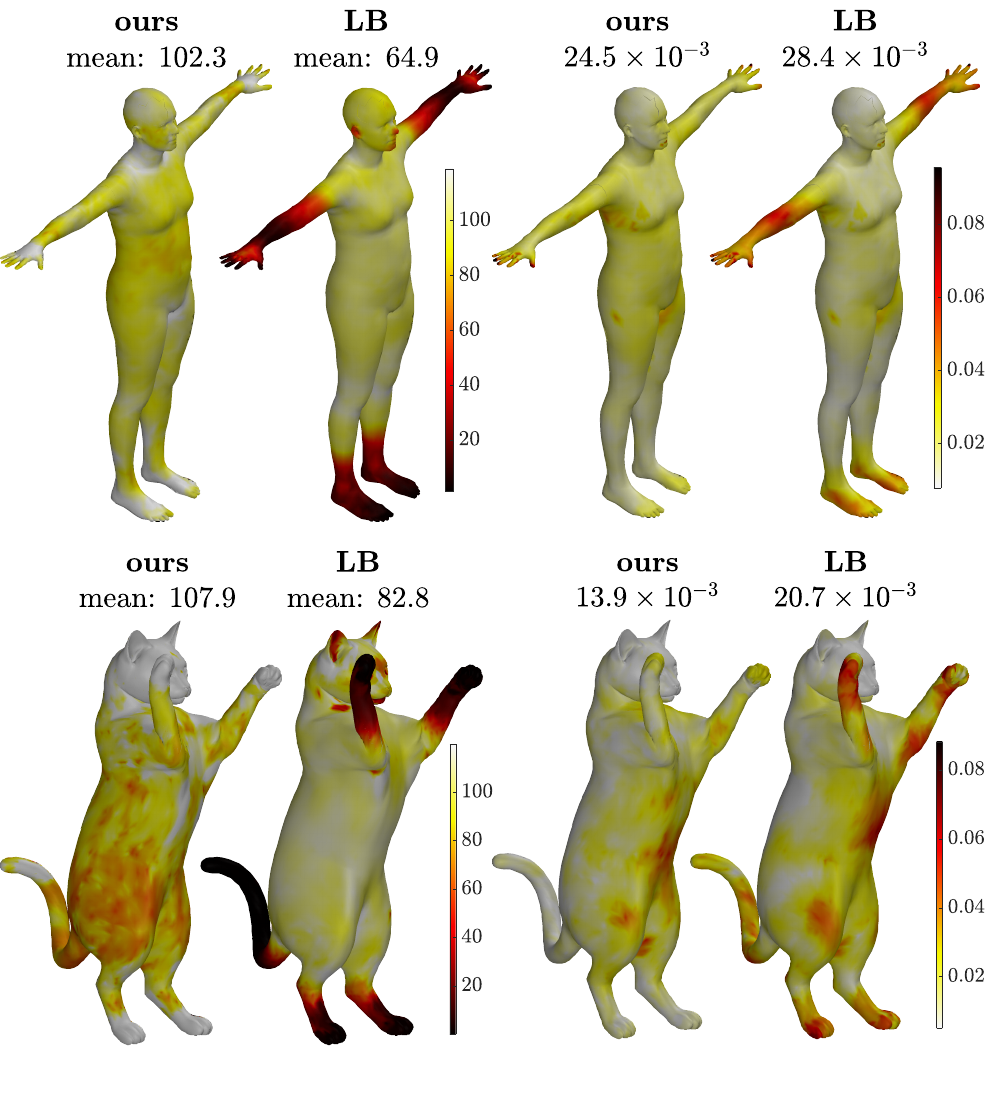}
    \caption{Spatial distribution of discrimination power of the basis and geodesic error. Comparison between \ours{} and LB. We visualize the average results across 40 human meshes from SHREC19 (top) and three cat meshes from TOSCA (bottom). Darker is worse in all cases. The similarity between distributions suggests a correlation between the error and the local quality of the embedding space.}
    \label{fig:localization}
\end{figure}

We define the \emph{discrimination power} as the capability of a basis to assign sufficiently different embeddings to different vertices. For each vertex $x\in V_{\M}$, we measure it with the following metric:
\begin{align}
    \discr(x) &= \frac{\norm{\emb(x)-\emb(y_{x})}_2}{\geo_{\M}\left(x,y_{x}\right)},\\
\text{ \CGnew{where} } \CGnew{y_{x}} &\CGnew{= \argmin_{z\in V_{\M}\setminus \{x\}} \left( \norm{\emb(z) - \emb(x)}_2 \right)}    
\end{align}   
Here, \CGnew{$y_{x}$} is the vertex having the closest embedding to $\emb(x)$ measured through Euclidean distance. The normalization makes the metric independent of vertex density and rewards geodesic proximity between $x$ and $y_{x}$. 
According to the continuous nature of the surface, it is reasonable to require that geodesically close points have similar embeddings.
For this reason, we penalize a vertex having as the nearest point in the embedding a vertex that is geodesically far. 

To assess the quality of \PCGAU{} embedding with respect to LB, we compare the distribution of their discrimination power over the mesh. Moreover, we investigate the relation between discrimination power and the average geodesic error, when the mesh is matched to another.
On the left of Figure~\ref{fig:localization}, we report the average distribution of $\discr(x)$ on meshes obtained across 40 human meshes from SHREC19 (top row) and three cat meshes from TOSCA (bottom row). On the right of Figure~\ref{fig:localization}, we visualize the average distribution of the geodesic error on the corresponding datasets matched using LB and \PCGAU{} as described in Section~\ref{ssec:exp_gt}. We observe that: 
\begin{enumerate} 
    \item LB has a much lower discrimination power on narrow extremities (arms, feet, paws, and tail), while \ours{} presents a uniform discrimination power similar to the best level achieved by LB.
    \item The error for LB is coherently localized in narrow extremities, while our diffuses uniformly on the surface.
\end{enumerate}
These observations support our claim that the errors in point-wise correspondence localize in the areas where the basis provides a less discriminative embedding. This result suggests that through \PCGAU{}, we can improve such correspondences by making the quality of the embedding space more uniform on the mesh.

\subsubsection{Locality preservation}
\label{ssec:locality_preservation}%
We define the following two metrics to quantitatively asses locality preservation of a given embedding space.
\paragraph*{Embedding/Geodesic Distance Correlation (EGDC)} 
For each vertex $x$, we select the set $S$ of $s\in \mathbb{N}$ vertices having the embedding closest to $\emb(x)$, and estimate the correlation ($\corr$) between the Euclidean distances of the embeddings and their geodesic distances from $x$:
\begin{equation}
    \egdc(x) = \corr\Big(\geo_{\M}(x,y), \norm{\emb(x)-\emb(y)}_2\Big)_{y\in S}
\end{equation}
This measure evaluates how much an embedding space preserves the neighborhood of $x$ in the geodesic sense. The higher $\egdc$, the better. In our experiments we assess $\egdc(x)$ using $s=80$ vertices.
\paragraph*{Mean Geodesic Distance (MGD)} 
For each vertex $x$, we define the set $R$ of the $t\in\mathbb{N}$ vertices having the lowest embedding distance to $\emb(x)$, and the set $\bar{R}$ of the $t$ vertices having the lowest geodesic distance from $x$. Then, we compute the ratio between the mean geodesic distance of the vertices in $R$ and $\bar{R}$:
\begin{equation}
    \mgd(x) = \dfrac{\avg_{y\in R}\left\{ \geo_{\M}(x,y) \right\} }{\avg_{z\in \bar{R}}\left\{ \geo_{\M}(x,z) \right\}}
\end{equation}
The more the embedding preserves the distance of the geodesic neighbor, the lower the value of $\mgd(x)$, which approaches 1. Therefore, the lower $\mgd(x)$ the better. In the evaluation we used $t = 10$.

\begin{figure}[ht]
    \centering
    \includegraphics[width=0.95\linewidth]{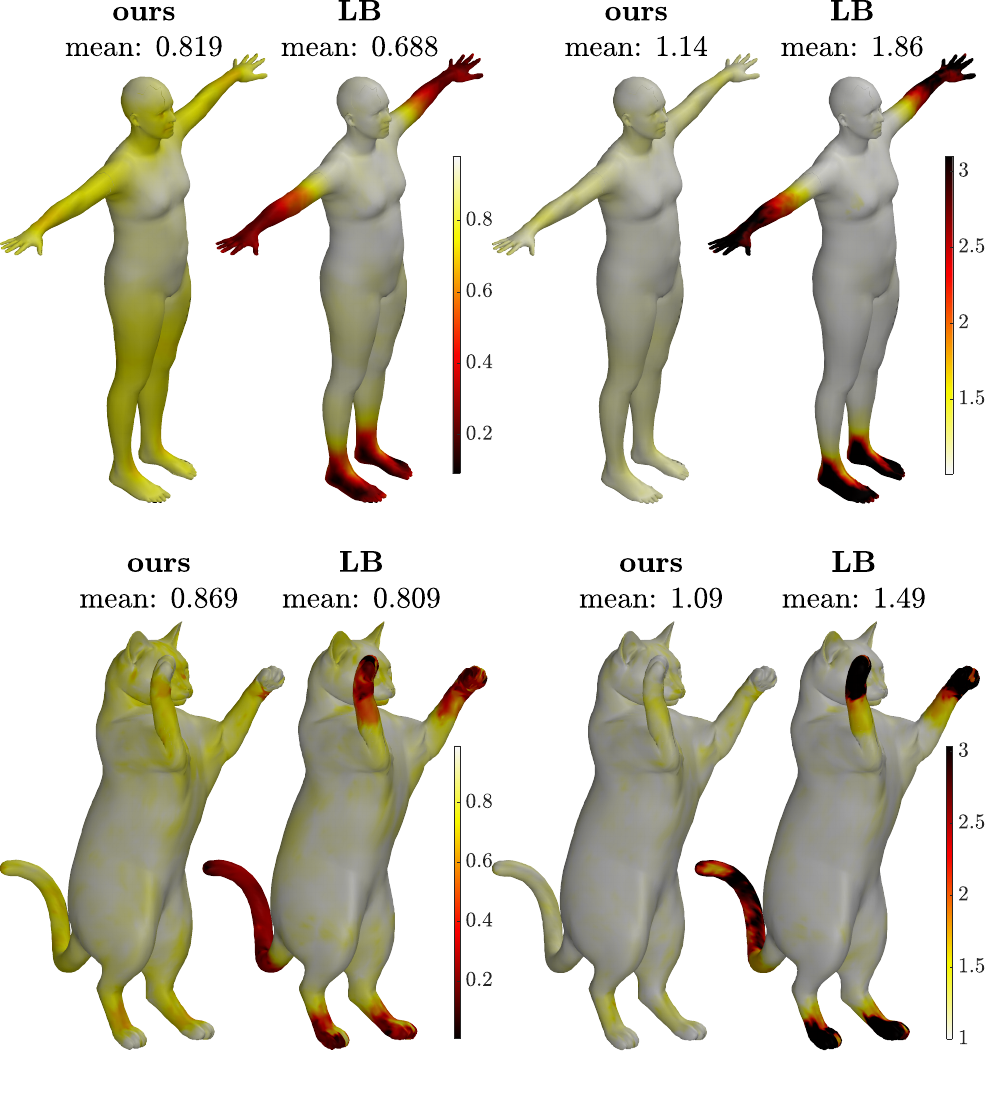}
    \caption{Spatial distribution of EGDC and MGD. Comparison between \ours{} and LB on average on human shapes from SHREC19 (top) and cats from TOSCA (bottom). Darker is worse in all cases. LB presents low locality preservation on the extremities (arms, feet, paws, and tail).}
    \label{fig:loc_preserv}
\end{figure}
We analyze the locality preservation of \PCGAU{}, compared to LB exploiting EGDC and MGD. We claim that point-wise embedding designed for shape matching should encode the neighbor information, preserving the relation of nearest and potentially similar points.
Figure~\ref{fig:loc_preserv} shows the value of these metrics averaged on the same shapes from SHREC19 (top) and TOSCA (bottom) involved in Figure~\ref{fig:localization}. The spatial distribution measured with EGDC and MGD for LB closely mimics the distribution of discrimination power (see Figure~\ref{fig:localization}, first column). The three similar distributions observed in these results support the choice of these measures to evaluate the quality of the embedding in different regions of the surfaces. 
In Table~\ref{tab:corr_nmgd}, we present the overall value for EGDC and MGD (computed as the average on the vertices of each mesh) for different datasets that we introduce in the next section. These results support our claim: a more even distribution of the energy of a basis, as exhibited by the one we propose, increases the overall quality of the embedding space.
\CGnew{\PCGAU{} inherits these properties from the Gaussian dictionary we propose. Providing a good approximation of the dictionary, and thus of the subspace $S$, and satisfying~\eqref{eq:pca_optimal}, our truncated basis $P_k$ implicitly preserves the uniform distribution owned by the Gaussian functions.}

\begin{table}[t]
\centering
\begin{tabular}{|l|cc|cc|}
\hline 
dataset & \multicolumn{2}{c|}{EGDC} & \multicolumn{2}{c|}{MGD}   \\ \hline
 & \textbf{ours}  & \textbf{LB}  & \textbf{ours} & \textbf{LB} \\ 
& $\times10^{-2}$ & $\times10^{-2}$  &    & \\ \hline \hline
FAUST   & \textbf{79,7}       & 73,8      & \textbf{1,11}    & 1,34    \\
MWG     & \textbf{83,8}       &  78,1     & \textbf{1,12}    & 1,36    \\
TOSCA   & \textbf{84,1}       &   79,0    & \textbf{1,12}    &   1,40  \\
SHREC19 & \textbf{81,9}       &   68,8    & \textbf{1,14}    &   1,86  \\ \hline
\end{tabular}
\caption{Overall values of EGDC and MGD averaged on the meshes of different datasets.}
\label{tab:corr_nmgd}
\end{table}

\section{Alternative Dictionaries}
\label{sec:dictionaries}
Even if the \PCGAU{} basis from~\cite{stag2022} already achieves the desired performance as a replacement for the LB in the functional map framework, the new \ourname{} framework is more general and applies to any dictionary $D$ constructed on a triangular mesh. In this section, we propose six alternative dictionaries, motivated both by the previous literature in geometry processing and by 
the properties analyzed in the previous section, that we showed to be relevant to improve the estimation of point-to-point correspondences. We briefly introduce each of them and refer the interested reader to the related literature for more details on each derivation.

\paragraph{Adaptative Gaussians (\emph{\textbf{ADAPT}})}
To define the Gaussian functions for \PCGAU{}, we fix the value of $\sigma$ and define for all the samples in $Q$ a Gaussian having the same covariance. Due to the diverse local structure of different regions on the surface, this choice can be suboptimal. For this reason, we define a procedure to vary the value of $\sigma$ on the surface to adapt the definition of each Gaussian to the local geometry in the neighborhood of its vertex.

We proceed as follows:
\begin{enumerate}
    \item We compute the error of reconstruction of the geometry ($\RECerr$) provided by an LB basis composed of 60 eigenfunctions that is:
\begin{align}
    \RECerr = \| X_{\M} - \Phi_\M \Phi_\M^{\dagger} X_{\M} \|_2,
\end{align}
where $X_{\M}$ are the vectors containing the 3D coordinates of the vertices of $\M$. $\RECerr$ is a function that encodes, for each vertex, the error in the reconstruction given by the selected LB basis. The larger the value of $\RECerr$, the larger the error in the reconstruction.
\item We normalize $\RECerr$ within the interval $[0, 1]$.
\item We define and adaptive $\widetilde{\sigma}(x)$ at each sample $x \in \M$ as:
\begin{equation}
    \widetilde{\sigma}(x) = \sigma (1-(\frac{2}{5} \RECerr )).
\end{equation}
\end{enumerate}  
This formula restricts the value of the adaptive $\widetilde{\sigma}$ in the interval $[\frac{3}{5}\sigma, \sigma]$. More specifically, $\widetilde{\sigma}(x)$ is smaller when the reconstruction error is larger. So we want to reduce the amplitude of the Gaussian in the regions where the representation provided by LB is worse. As for all our experiments, we set $\sigma = 0.05$ and refer to this dictionary as \textbf{ADAPT}. This is not the only viable solution to produce an adaptive dictionary. In our procedure, we have tested different definitions varying the formula to compute $\RECerr$, the size of the LB basis, and the value of $\frac{2}{5}$ in the last formula. A comprehensive exploration of the adaptive dictionary in our pipeline is out of the scope of our analysis, and we leave this task to future exploration.

\paragraph{Heat Gaussians (\emph{\textbf{HEAT}})}
We consider the heat diffusion and, in particular, the heat kernel, which encodes for each pair of points $x,y \in \M$ the amount of heat transferred from $x$ to $y$ in a time $t$ given a unit heat source in $x$.
This quantity is a function of the LB eigenvalues and eigenfunctions, and thus, it is fast to compute as described in~\cite{sun2009concise}:
\begin{align}
    h_t(x,y) &= \sum_{\ell = 1}^{k} e^{-t\lambda_{\ell}}\phi_{\ell}(x)\phi_{\ell}(y).
\end{align}
With this equation, we can define a dictionary $D$ where the $j$-th atom, $\forall j \in Q$, is equal to $h_t(j,*)$. $h_t(j,*)$ is a function that $\forall y \in \M$ has value $h_t(j,y)$, i.e. the amount of heat transferred from $j$ to $y$ in time $t$ starting from a unit heat source in $j$ at time $0$. We set $t=1\times10^{-2}$ to obtain a heat diffusion with an amplitude similar to the one of the Gaussians in \PCGAU{}. Being fully determined by the LB, the heat kernel is an isometric-invariant quantity and is strictly related to the geodesic distance~\cite{geodesicheat}. The dictionary we build from the heat kernel, namely \textbf{HEAT}, inherits all these properties and the ones given by the uniform sampling of the $Q$ subset of vertices we use to define it.

\paragraph{Spectral Gaussians (\emph{\textbf{SPEC}})}
We take inspiration from~\cite{boscaini15} and~\cite{melzi19step} for a different definition of the dictionary of Gaussian functions. 
We define a Gaussian filter on the frequency domain and interpret them as a set of coefficients to linearly combine the LB eigenfunctions obtaining a Gaussian function in the space domain (i.e. on the surface).
Following~\cite{melzi19step}, we can define these coefficients as:
\begin{align}
    \hat{g}(\lambda_\ell) &= e^{-\alpha\lambda_\ell}, \ \forall \ell\in \{1, \ldots , k\},
\end{align}
where $\alpha \in \R$ controls the amplitude of the Gaussians (we fix the parameter $\alpha=1\times10^{-4}$).
Then, to reconstruct the value $g_{\alpha,j}(y), \forall y\in \M$ of the Gaussian centered at $j\in Q \subseteq \M$ we compute:
\begin{align}
    g_{\alpha,j}(y) &= \sum_{\ell = 1}^{k}  \hat{g}(\lambda_\ell) \phi_{\ell}(j)\phi_{\ell}(y).
\end{align}
We refer to this dictionary in which atoms correspond to $g_{\alpha,j} \forall j \in Q$, as \textbf{SPEC}. Being linear combinations of the LB eigenfunctions, the atoms of \textbf{SPEC} are invariant to isometries. Furthermore, the generated basis respects the same prior given by the even distribution of the samples in $Q$.

\paragraph{Wave Kernel Signature (\emph{\textbf{WKS}})}
As described in Section~\ref{sec:background}, it is possible to optimize a functional map by imposing functional constraints to preserve some point-wise descriptor. In our experiments, we adopt the method NO17~\cite{nogneng17}, which exploits the \emph{wave kernel signature}~\cite{aubry2011wave} (WKS) to produce such constraints. WKS is a multiscale descriptor based on the solution of the wave equation on a non-Euclidean domain. As described in~\cite{nogneng17}, as for the heat diffusion, it is possible to approximate the wave diffusion through the eigenfunctions and the eigenvalues of the LB. This approximation is fast to compute and makes this descriptor invariant to isometries. For these reasons, it is reasonable to consider the different scales of WKS as atoms of a dictionary. We compute WKS with the same parameters from~\cite{nogneng17}, and we refer to~\cite{aubry2011wave} for more details. With this choice, we force the orthonormal bases produced by our method to improve the representation of the functions which define the optimization constraints. We refer to the dictionary composed by only the WKS as \textbf{WKS}.
Additionally, we consider a dictionary, which contains both Gaussian functions, computed as in Section~\ref{sec:gaussian_dictionary}, and the wave kernel signature. We refer to this dictionary as \textbf{WKS+\PCGAU{}}. Note that in this case we are considering a dictionary composed by functions from different families. This is perfectly possible, as our framework does not pose any kind of constraint on the composition of the initial dictionary.

\paragraph{Wavelets (\emph{\textbf{WAVE}})}
Taking inspiration from the signal processing literature, we replace Gaussians by considering wavelet functions centered in the same set of sampled vertices $Q \subset V_{\M}$ as atoms of our dictionary. The main motivation is that wavelets should provide a more informative representation of the region around each sample. To define the wavelets, we adopt the procedure proposed in~\cite{kirgo2021} for its efficiency and relation with heat diffusion and we exploit the code provided in~\cite{kirgo2021}. We refer to the obtained dictionary as \textbf{WAVE}.

\begin{table*}[t]
    \centering
\begin{tabular}{|c|l|c c c c c c c | c |}
\hline

 & dataset $\downarrow$ & \PCGAU{} &  \textbf{ADAPT} & \textbf{HEAT} & \textbf{SPEC} & \textbf{WKS} & \textbf{WKS}+\PCGAU{} & \textbf{WAVE} & \textbf{LB} \\ \hline \hline
 \multirow{7}{*}{\rotatebox{90}{GT}} & FAUST 1:1    & \tabblue{\textbf{8,4}} & \tabgreen{\textbf{7,8}} & 15,0 & 19,6 & 42,6 & \tabred{\textbf{7,2}} & 27,1 & 13,6 \\
 & FAUST     & \tabblue{\textbf{15,7}} & \tabgreen{\textbf{15,3}} & 20,2 & 23,6 & 56,2 & \tabred{\textbf{14,5}} & 28,1 & 19,7 \\
  & MWG       & \tabblue{\textbf{20,8}} & \tabgreen{\textbf{20,1}} & 24,7 & 31,7 & 94,2 & \tabred{\textbf{18,4}} & 30,4 & 24,7 \\
 & MWG iso   & \tabblue{\textbf{13,5}} & \tabgreen{\textbf{13,1}} & 17,3 & 25,8 & 99,0 & \tabred{\textbf{11,7}} & 24,7 & 17,3 \\
& TOSCA     & \tabgreen{\textbf{6,3}} & \tabgreen{\textbf{6,3}} & 9,8 & 13,0 & 92,1 & \tabred{\textbf{5,8}} & 16,6 & 9,9 \\
& TOSCA cat     & \tabblue{\textbf{12,8}} & \tabgreen{\textbf{12,5}} & 19,7 & 22,9 & 41,0 & \tabred{\textbf{12,0}} & 21,3 & 19,8 \\
& SHREC19   & \tabgreen{\textbf{24,5}} & \tabred{\textbf{24,0}} & 28,6 & 31,2 & 88,6 & 36,7 & 34,3 &  \tabblue{\textbf{28,4}} \\  \hline \hline

 \multirow{7}{*}{\rotatebox{90}{NO17}} & FAUST 1:1  & \tabblue{\textbf{16,4}} & \tabgreen{\textbf{15,9}} & 28,3 & 313,8 & 451,6 & \tabred{\textbf{14,9}} & 38,4 & 29,9 \\
 & FAUST      & \tabgreen{\textbf{28,0}} & \tabblue{\textbf{28,8}} & 29,8 & 346,4 & 387,6 & \tabred{\textbf{26,2}} & 41,3 & 30,6 \\
 & MWG       & 61,3 & 61,2 & \tabgreen{\textbf{59,7}} & 295,6 & 473,1 & \tabred{\textbf{55,2}} & 87,4 & \tabblue{\textbf{60,6}} \\
 & MWG iso   & \tabgreen{\textbf{26,0}} & 27,7 & 27,9 & 202,1 & 471,6 & \tabred{\textbf{18,8}} & 44,8 & \tabblue{\textbf{27,5}} \\
& TOSCA      & \tabred{\textbf{12,9}} & \tabgreen{\textbf{13,1}} & 18,1 & 409,8 & 462,4 & \tabblue{\textbf{13,9}} & 46,7 & 19,9 \\
& TOSCA cat     & \tabgreen{\textbf{29,2}} & \tabblue{\textbf{29,8}} & 35,4 & 295,0 & 374,4 & \tabred{\textbf{28,6}} & 60,6 & 34,7 \\
& SHREC19    & \tabred{\textbf{43,3}} & \tabgreen{\textbf{44,5}} & 66,6 & 404,0 & 486,0 & \tabblue{\textbf{47,8}} & 118,7 & 65,9\\  \hline \hline

 \multirow{7}{*}{\rotatebox{90}{ZoomOut}} &
FAUST 1:1    & \tabgreen{\textbf{13,8}} & \tabred{\textbf{13,5}} & 24,0 & 36,3 & 235,2 & \tabblue{\textbf{26,3}} & 33,7 & 31,3 \\
& FAUST      & \tabblue{\textbf{24,6}} & \tabgreen{\textbf{24,4}} & 26,5 & 30,5 & 241,8 & \tabred{\textbf{23,1}} & 33,5 & 26,0 \\
 & MWG        & \tabgreen{\textbf{49,6}} & \tabred{\textbf{49,5}} & 70,0 & 72,6 & 311,4 & \tabblue{\textbf{59,7}} & 85,2 & 67,9 \\
 & MWG iso    & \tabred{\textbf{18,6}} & \tabgreen{\textbf{19,9}} & 30,7 & 34,6 & 281,7 & \tabblue{\textbf{22,9}} & 33,2 & 27,4 \\
& TOSCA      & \tabred{\textbf{10,9}} & \tabgreen{\textbf{11,1}} & 18,8 & 19,7 & 215,4 & \tabblue{\textbf{12,5}} & 22,3 & 18,0 \\
& TOSCA cat     & \tabgreen{\textbf{26,1}} & \tabred{\textbf{25,5}} & 35,1 & 36,3 & 164,8 & \tabblue{\textbf{28,8}} & 39,7 & 35,7 \\
& SHREC19    & \tabgreen{\textbf{35,7}} & \tabred{\textbf{35,1}} & 38,0 & 44,9 &  312,9 & 39,5 & 42,8 & \tabblue{\textbf{39,2}} \\  \hline 
\end{tabular}
\caption{\CGnew{Comparison on the Average Geodesic Error ($\age$) of point-wise maps estimated from a functional map and representing the functional spaces with LB or with the proposed approach considering seven different input dictionaries (one for each column). We divide the rows of the table into three parts, which correspond to different computations of the functional maps, from top to bottom: computed from ground-truth correspondence, estimated with~\cite{nogneng17}, and estimated with ZoomOut~\cite{melzi2019zoomout}. In each row, we consider a different test set. For each dataset (row), we depict the $\age$ value of the best, second best and third best, respectively, in \tabred{\textbf{red}}, \tabgreen{\textbf{green}} and \tabblue{\textbf{blue}} to highlight the distribution of the more accurate solutions. All the results reported in the table must be multiplied by a factor equal to $ 10^{-3}$.} 
} 
\label{tab:dictionaries}
\end{table*}

\section{Experimental Evaluation}
\label{sec:exp}%
The purpose of \ourname{} is to improve the quality of the estimated point-wise map when employed in a functional maps pipeline for shape-matching. Therefore, the primary evaluation criteria consists in assessing the accuracy of point-wise maps between pairs of meshes. In the following, we compare the results between \ourname{} and LB in different settings \CGnew{and for different choices of the dictionary $D$}. Since \ourname{} and LB can be interchangeably used in the functional maps framework, it is easy to set up the same pipeline and compare the final results obtained in the same conditions.

\paragraph*{Metrics} We evaluate the overall accuracy of a point-wise map $\myT:V_{\N}\to V_{\M}$ by the average geodesic error: 
\begin{equation}
    \age(\myT) = \avg_{x\in V_{\N}} \{e(x)\},
\end{equation} 
where $e(x) = \geo_{\M}(\myT(x), \gtT(x))$ is computed with respect to the ground-truth correspondence $\gtT$ provided by the dataset for evaluation. 
Since our goal is to improve the performance compared to LB, we can also consider the Relative Error of \ourname{} with respect to LB to make the accuracy gain more explicit:
\begin{equation}
    \re(\myT_{\text{ours}}, \myT_{\text{LB}}) = \frac{\age(\myT_{\text{ours}})-\age(\myT_{\text{LB}})}{\age(\myT_{\text{LB}})}.
\end{equation} 
To evaluate the overall performance on a dataset, we can compute the mean of $\age$ and $\re$ over all the pairs considered in the dataset.
A negative value of MRE (Mean Relative Error) indicates that, on average, \ourname{} is performing better than LB on the given dataset. Note that MRE does not coincide with the relative difference of the mean AGE, and the assessment provided by $\re$ is less dependent on the absolute complexity of the pairs considered.

\paragraph*{Datasets}
\CGnew{As test sets, we generate} random pairs of meshes taken from standard datasets. We normalize all the meshes to the unitary area, which is a standard \CGnew{unsupervised} practice to compare the errors among different datasets. 
\begin{itemize}
\item \CGnew{\textbf{FAUST 1:1}} \cite{FAUST} is a dataset composed of 10 human subjects in 10 poses each. \CGnew{All the shapes in this dataset share a fixed template that provides a 1:1 correspondence among their vertices. We additionally consider a version of this dataset, namely \textbf{FAUST}, in which all the shapes have been independently remeshed to avoid the implicit use of additional knowledge about the meshes.} We used 200 random pairs \CGnew{for both \textbf{FAUST 1:1} and FAUST}.
\item \textbf{MWG} contains 25 meshes representing a man, a woman, \CGnew{or} a gorilla, in different poses and with different connectivities. We indicate with \textbf{MWG iso} the dataset containing only the man and woman meshes. We used 200 random pairs form MWG and 120 from MWG iso. 
\item \textbf{TOSCA} \cite{TOSCA} is a benchmark of 3D shapes divided in different classes. Meshes are high resolution ($\sim$50K vertices), isometric, and with vertices in 1:1 correspondence in each class. We limit our quantitative analysis to 20 pairs involving five meshes from the class Michael (one of the human subjects) in different poses. \CGnew{Moreover, in quantitative and qualitative evaluations, we consider pairs composed only of different poses from the cat class. We refer to this test set as \textbf{TOSCA cat}.} 
\item \textbf{SHREC19} \cite{SHREC19} contains 40 human bodies. These meshes largely differ in the number of vertices, type of tessellation, and model style, making it a challenging dataset. We used 200 random pairs.
\end{itemize}
In MWG we only have a sparse set of ground truth correspondences (around 1K correspondences). \CGnew{We have a complete ground truth correspondence for all the other datasets.} In other words, there is always one point on the target that corresponds to each vertex of the source shape.

\paragraph*{Implementation}
We implemented the procedure to build \ourname{} and the entire experimental setting in \textsc{Matlab}. We adopt the code available online for ZoomOut~\cite{melzi2019zoomout} and~\cite{nogneng17} provided by the authors. In all the following experiments, we fix the number of atoms $k = 60$ \CGnew{and the specific parameters required by each dictionary selected as described in the previous Sections.} 


\subsection{Functional maps computed from ground-truth correspondence}
\label{ssec:exp_gt}%
\begin{figure*}[!hbtp]
    \centering
    \includegraphics[width =\linewidth]{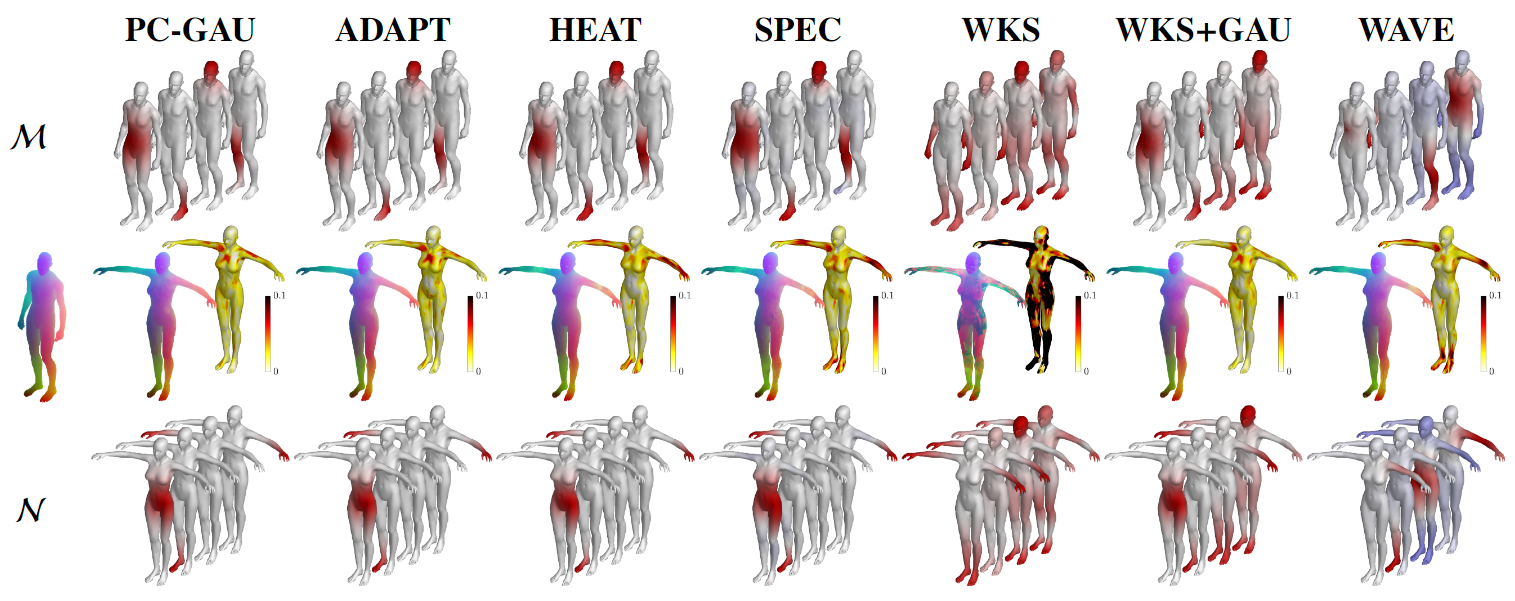}
    \caption{\CGnew{An example of four atoms for each considered dictionary, represented on a random pair $\mathcal{M}$ (top), $\mathcal{N}$ (bottom) taken from MWG. In the middle row, obtained point-wise maps and the corresponding geodesic error are shown. The map is computed from a ground-truth $C$. Note that, even if atoms with the same indices between $\mathcal{M}$ and $\mathcal{N}$ are visualized, their localization on the mesh is not the same. This is because the points in $Q_{\mathcal{M}}$ and $Q_{\mathcal{N}}$ are selected independently.}}
    \label{fig:pMaps_all_dicts}
\end{figure*}

To evaluate our basis independently of the quality of \CGnew{the estimation of the functional map}, we start by computing $C$ from the ground truth correspondence provided by each dataset. In this case, $C$ is computed as in Equation~\ref{eq:CfromGT} and can be considered the best possible functional correspondence \CGnew{in the given bases} $\Phi_{\M}$ and $\Phi_{\N}$. For the conversion to point-wise map, we adopt the original algorithm proposed in~\cite{ovsjanikov2012functional} and presented in Section~\ref{ssec:p2p_conv}. \CGnew{The first seven rows of Table~\ref{tab:dictionaries}} show the results on the different datasets. \CGnew{In this Table, for each dataset (row), we depict the $\age$ value of the best, second best and third best, respectively, in red, green and blue to highlight the distribution of the more accurate solutions.} We observe that \ourname{} provides substantially more accurate conversions. \CGnew{Among the different dictionaries, \textbf{WKS+\PCGAU{}} achieves the best result in six of the seven test sets failing only on the SHREC19 dataset. Even if not so accurate, \textbf{\PCGAU{}} and \textbf{ADAPT} are competitive and outperform the LB. The other four solutions are pretty far from the best results.} 
\CGnew{In Figure~\ref{fig:pMaps_all_dicts}, we show a qualitative comparison for a random pair from the MWG dataset among the different dictionaries in shape matching estimated through the functional framework from the ground truth correspondence. For each dictionary, we show some of the atoms that compose the dictionary on the two shapes (respectively on top and bottom). In the middle, we report the color coding of the matching and the geodesic error represented by the colors. White points have a $0$ error, while the darker the color, the larger the error).}

\subsection{Functional maps estimated with product preservation}
\label{ssec:exp_no17}%
In a real-world setting, we cannot compute $C$ from the ground-truth correspondence, but we need to estimate it. \CGnew{The second block of seven rows in Table~\ref{tab:dictionaries}} presents the results when $C$ is estimated using product preservation as functional constraints~\cite{nogneng17}\CGnew{, namely NO17}. 
In these tests, we used functional constraints based on six landmarks (\CGnew{placed on the extremities like feet, hands, head, and chest for human}) and the WKS 
descriptor~\cite{aubry2011wave}. 
\CGnew{\ourname{} produces better results than LB. Only on MWG and MWG iso LB obtain the third best score. The comparison among the different dictionaries is similar to the ground-truth case, with \textbf{WKS+\PCGAU{}} being the best one, but presenting some difficulties in targeting SHREC19. The difference among \textbf{WKS+\PCGAU{}}, \textbf{\PCGAU{}} and \textbf{WKS} reveals that having WKS in the dictionary is useful but only if supported by an evenly distributed set of atoms like the Gaussians from \textbf{\PCGAU{}}.}

\subsection{Functional maps estimated with ZoomOut}
\label{ssec:exp_zoomout}%
Finally, we tested our basis with the iterative refinement introduced in~\cite{melzi2019zoomout}. The last block of rows in \CGnew{Table~\ref{tab:dictionaries}} contains these results. This technique heavily takes advantage of the conversions from functional to point-wise maps and vice-versa. The results show that \ourname{} outperforms LB. In these experiments, we used an initial map of size $k_{\text{ini}} = 16$, estimated with~\cite{nogneng17} with the same constraints as in Section~\ref{ssec:exp_no17}. We then adopt the same parameters from~\cite{melzi2019zoomout}, increasing the size of the functional map to $k_\text{final}=60$ with step equal to 2. The great performance achieved by our basis, when combined with ZoomOut, confirms that \ourname{} is better suited to represent point-wise correspondences in the functional paradigm. Moreover, with ZoomOut, our method seems more stable also in the presence of non-isometries. \CGnew{The comparison among the different dictionaries behaves similarly to the other cases even if with ZoomOut \textbf{\PCGAU{}} and \textbf{ADAPT} perform better than \textbf{WKS+\PCGAU{}}, with a little gap which favours \textbf{ADAPT}.}

\CGnew{A comprehensive assessment of the results in Table~\ref{tab:dictionaries} highlights that \textbf{\PCGAU{}}, \textbf{ADAPT} and \textbf{WKS+\PCGAU{}} are not only comparable to LB's performance but even superior. The other dictionaries, which do not involve Gaussians defined through geodesic distances, give rise to less accurate correspondences. \textbf{WKS}, the only solution not based on the sampling strategy and the generation of localized functions, is the worst dictionary and confirms the goodness of the sampling strategy adopted in~\cite{stag2022}. Moreover, the atoms of \textbf{WKS} do not discern the symmetric points because, being entirely intrinsic, they are quasi-symmetric. Due to this property, the representation obtained from \textbf{WKS} with our pipeline can give rise to some errors in symmetric shapes. On the contrary, \textbf{HEAT} and \textbf{SPEC}, which arise from the sampling strategy and the definition of local Gaussians, do not perform well compared to the solutions that exploit the geodesic distances. The spectral Gaussians and the heat kernel are more sensitive to the slight variation in the intrinsic geometry. They do not provide a point representation that permits the correct matching of the corresponding points. The \textbf{WAVE} dictionary, which produces less precise maps, does not adopt the geodesic distances to define its atoms. Furthermore, for the \textbf{WAVE} solution, we select the efficient and recent solution from~\cite{kirgo2021}, maintaining the parameters provided by the authors. A more accurate analysis to select the parameters that define the wave functions is necessary to build a more informative dictionary of waves.}

\subsection{\CGnew{Direct comparison between \PCGAU{} and LB}}
\CGnew{The previous quantitative comparison does not establish a clear champion among the candidate dictionaries, even if it clearly shows that \textbf{\PCGAU{}}, \textbf{ADAPT} and \textbf{WKS+\PCGAU{}} well exceed the accuracy of LB.
 \textbf{WKS+\PCGAU{}} is probably the best candidate with both ground-truth and NO17 estimation, but it does not perform well with the ZoomOut refinement that is currently the most selected method to estimate functional map.
 \textbf{ADAPT} works well with ZoomOut, but its performance is similar to the one of \textbf{\PCGAU{}}, which does not require additional computation to produce an adaptive $\sigma$.
For these reasons, we consider \textbf{\PCGAU{}} as our representative choice and leave further exploration to strengthen \ourname{} by defining more appropriate dictionaries as an intriguing future direction.}

\CGnew{Focusing on \textbf{\PCGAU{}}, we perform a direct comparison to LB that we report in Table~\ref{tab:age}, adding the $\re$ for each test set. We consider again the same three estimations of the functional maps, now organized as three blocks on the columns of Table~\ref{tab:age}.
In the first three columns, we observe that \textbf{\PCGAU{}} provides substantially more accurate conversions with the ground-truth functional maps.} Results in terms of MRE are consistent with the average geodesic error. The right part of Figure~\ref{fig:localization} shows the average spatial distribution of the error on the considered pairs from SHREC19 and cats from TOSCA.

\begin{figure}
    \centering
    \subfloat[FAUST]{\includegraphics[width=0.48\linewidth]{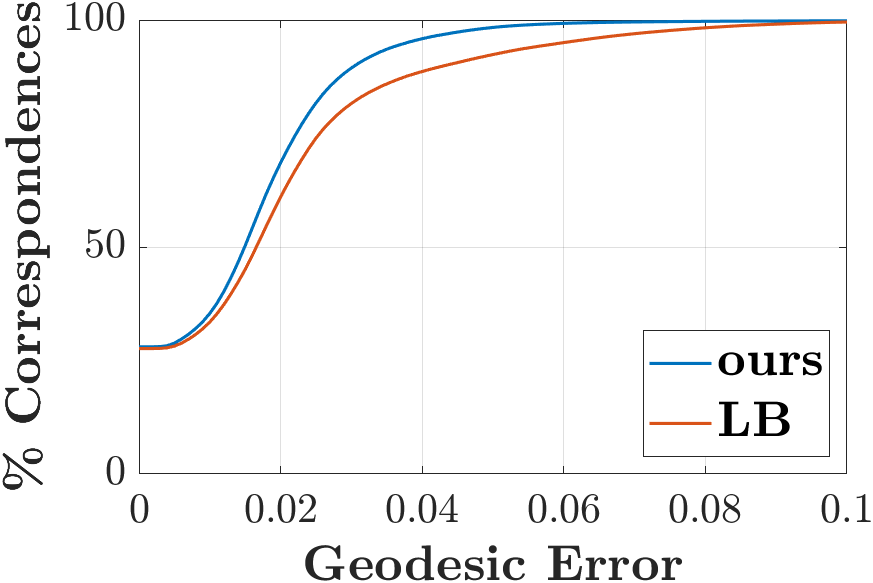} \label{sfig:curves_gt:faust}} \hfill
    \subfloat[TOSCA]{\includegraphics[width=0.48\linewidth]{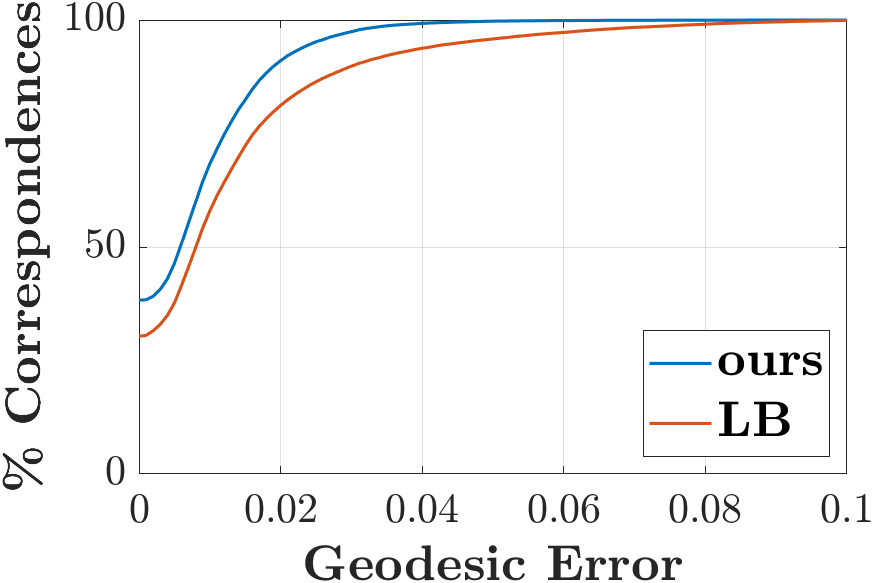}\label{sfig:curves_gt:tosca}} \hfill
    \subfloat[MWG]{\includegraphics[width=0.48\linewidth]{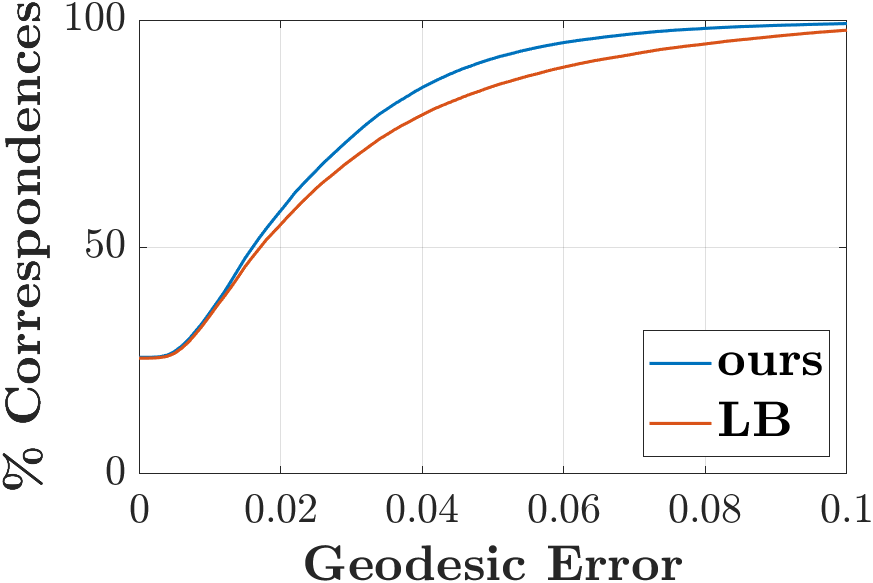} \label{sfig:curves_gt:mwg}} \hfill
    \subfloat[SHREC19]{\includegraphics[width=0.48\linewidth]{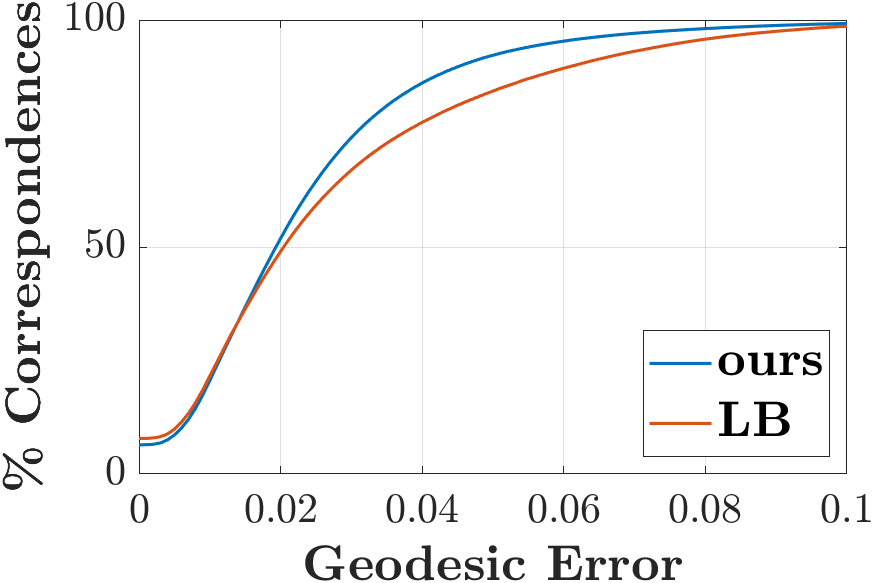} \label{sfig:curves_gt:shrec19}}
    \caption[Curves of Cumulative Geodesic Error for ground-truth $C$]{Curves of the cumulative geodesic error of point-wise maps converted from ground-truth $C$ for the four datasets. For each curve point, the $y$-axis represents the percentage of vertices with a geodesic error lower than the error threshold corresponding to the $x$-axis: the higher the curve, the more precise the estimated correspondence.}
    \label{fig:curves_gt}
\end{figure}

Following the protocol proposed in~\cite{kim2011blended}, Figure~\ref{fig:curves_gt} shows the curves of the cumulative geodesic error for each dataset (more details in the caption). Note that the percentage of vertices with a low geodesic error is similar between \textbf{\PCGAU{}} and LB, but \textbf{\PCGAU{}} has a considerably lower percentage of vertices with a large error. This result is consistent with our analysis: the embedding space of LB is not poor in general but only in specific areas of the mesh. The vertices in the weakly-represented regions have a large geodesic error when using LB, while the quality of point-wise maps is similar between \textbf{\PCGAU{}} and LB in the well-covered areas.

\CGnew{With NO17~\cite{nogneng17} (columns 4,5, and 6),} \textbf{\PCGAU{}} still gets better results than LB, except in MWG. Removing the gorilla meshes from the dataset brings the advantage back (see row MWG iso), suggesting that \textbf{\PCGAU{}} is more susceptible to the error raised by non-isometries when we compute the $C$ adopting NO17 and WKS as descriptors. In all the cases, even for all the tests with the Zoomout refinement, \textbf{MRE}, which is less sensitive to the single pair challenges, is always negative, confirming that our basis outperforms LB.

\begin{table*}[ht]
    \centering
\begin{tabular}{|l| ccc | ccc | ccc|}
\hline
dataset   & \multicolumn{3}{c|}{GT} & \multicolumn{3}{c|}{NO17} & \multicolumn{3}{c|}{ZoomOut} \\ \hline
\textbf{} & \textbf{ours}                 & \textbf{LB}                   & \textbf{MRE}                  & \textbf{ours}                 & \textbf{LB}                   & \textbf{MRE}                  & \textbf{ours}                 & \textbf{LB}                   & \textbf{MRE}                  \\
          & \footnotesize$\times 10^{-3}$ & \footnotesize$\times 10^{-3}$ & \footnotesize$\times 10^{-2}$ & \footnotesize$\times 10^{-3}$ & \footnotesize$\times 10^{-3}$ & \footnotesize$\times 10^{-2}$ & \footnotesize$\times 10^{-3}$ & \footnotesize$\times 10^{-3}$ & \footnotesize$\times 10^{-2}$ \\ \hline \hline
\CGnew{FAUST 1:1} & \textbf{8,4} & 13,6 & \textbf{-37,8} 
            & \textbf{16,4} & 29,9 & \textbf{-33,0}
            & \textbf{13,8} & 31,3 & \textbf{-43,5} \\ 
FAUST     & \textbf{15,7}                 & 19,7                          & \textbf{-20,3}                & \textbf{28,0}                 & 30,6                          & \textbf{-4,9}                 & \textbf{24,6}                 & 26,0                          & \textbf{-5,4}                 \\
MWG       & \textbf{20,8}                 & 24,7                          & \textbf{-19,8}                & {61,3}                        & \textbf{60,6}                 & \textbf{-4,9}                 & \textbf{49,6}                 & 67,9                          & \textbf{-27,6}                \\
MWG iso   & \textbf{13,5}                 & 17,3                          & \textbf{-25,6}                & \textbf{26,0}                 & 27,5                          & \textbf{-9,4}                & \textbf{18,6}                 & 27,4                          & \textbf{-29,7}                \\
TOSCA     & \textbf{6,3}                  & 9,9                           & \textbf{-39,1}                & \textbf{12,9}                 & 19,9                          & \textbf{-39,0}                & \textbf{10,9}                 & 18,0                          & \textbf{-43,7}                \\
\CGnew{TOSCA cat} & \textbf{12,8} & 19,8 & \textbf{-35,9}
            & \textbf{29,2} & 34,7 & \textbf{-17,5}
            & \textbf{26,1} & 35,7 & \textbf{-28,8} \\ 
SHREC19   & \textbf{24,5}                 & 28,4                          & \textbf{-14,0}                & \textbf{43,3}                 & 65,9                          & \textbf{-15,4}                & \textbf{35,7}                 & 39,2                          & \textbf{-7,1}                \\ \hline
\end{tabular}
\caption{Average Geodesic Error (both absolute and relative) of point-wise maps converted from a $C$ (from left to right): computed from ground-truth correspondence, estimated with \cite{nogneng17}, and estimated with ZoomOut~\cite{melzi2019zoomout}. 
}
\label{tab:age}
\end{table*}

\begin{figure}[hbtp]
    \centering
    \subfloat[discrimination power]{\includegraphics[width=0.48\linewidth]{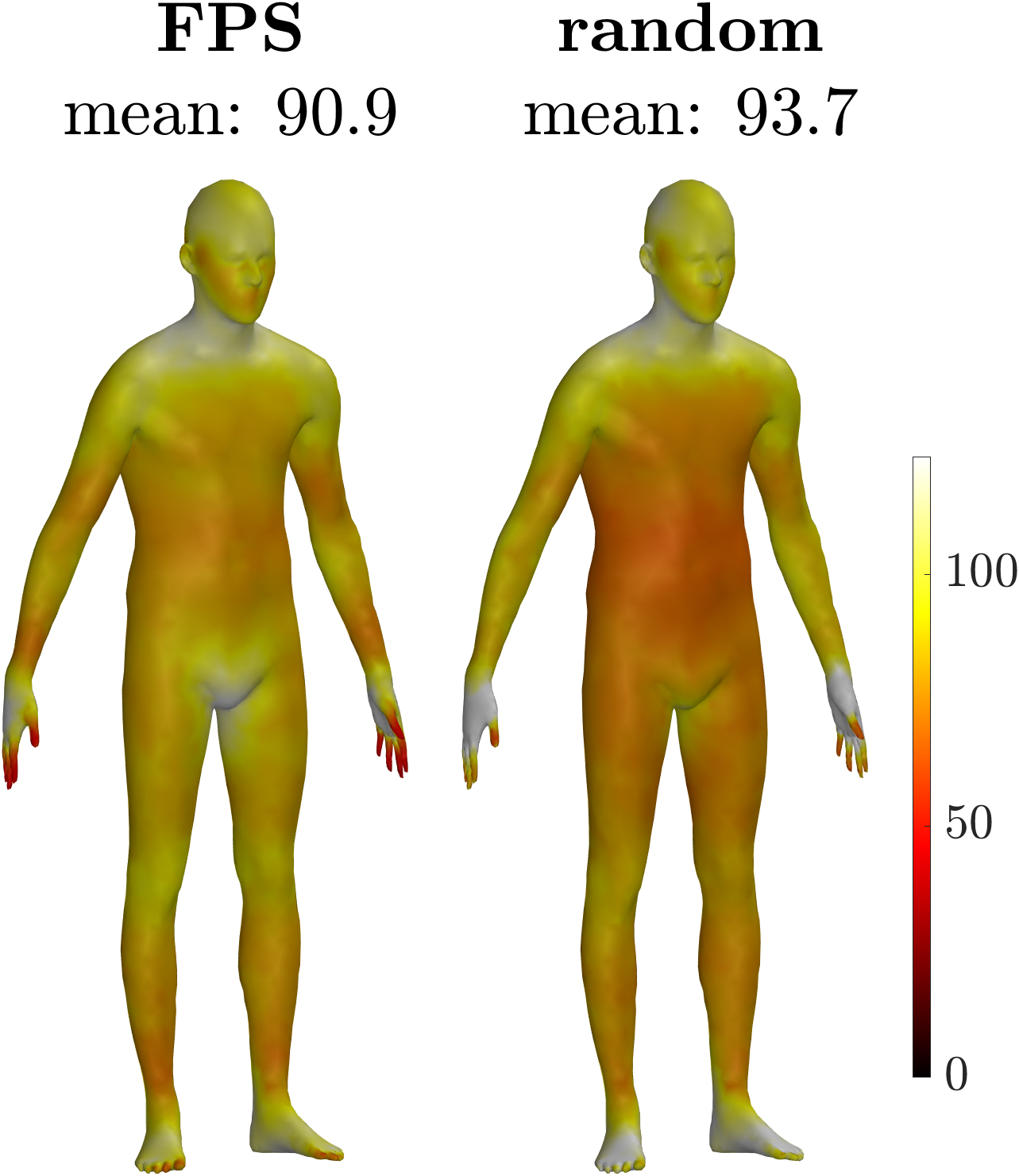} \label{sfig:random_sampling:basis}} \hfill
    \subfloat[geodesic error]{\includegraphics[width=0.48\linewidth]{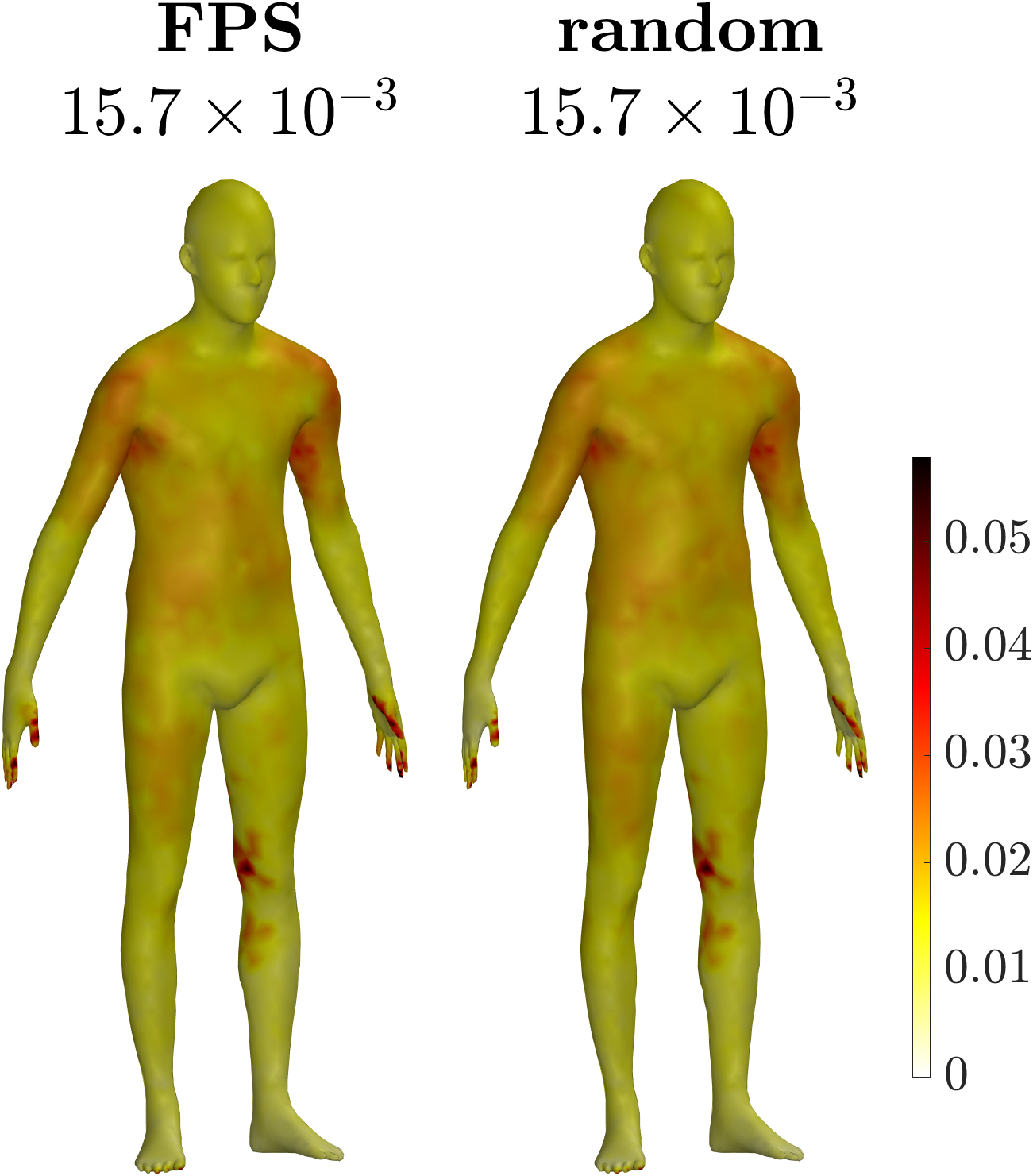} \label{sfig:random_sampling:error}}
    \caption{Comparison between random sampling and FPS for the selection of $Q$ on the FAUST dataset. \protect\subref{sfig:random_sampling:basis} shows the average distribution of the discrimination power of the obtained bases. \protect\subref{sfig:random_sampling:error} shows the distribution and the overall value of the geodesic error for point-wise maps obtained from the two bases. We compute $C$ exploiting the provided ground-truth correspondence.}
    \label{fig:random_sampling}
\end{figure}

\paragraph*{Random sampling of $Q$}
The quality of our basis does not depend on the exact composition of the subset of points where the Gaussians are initially computed, as far as the scattering of vertices in $Q$ is sufficiently uniform on the mesh surface. To verify this claim, we build \textbf{\PCGAU{}} by randomly selecting $Q$ (without replacement) instead of using Farthest Point Sampling~\cite{moenning03FPS}, as described in \ref{sec:gaussian_dictionary}. Figure~\ref{fig:random_sampling} compares the bases obtained with the two methods regarding both discrimination power and accuracy of the final point-wise map. Discrimination power shows a similar distribution between the two, with \textbf{\PCGAU{}} computed from FPS being slightly more uniform over the surface (look at the difference in value between chest and feet, for instance). This difference, however, does not impact the point-wise maps obtained, which show almost identical geodesic errors. Therefore, \textbf{\PCGAU{}} is robust to the algorithm for selecting $Q$. We still prefer using FPS over random sampling, because this makes the basis less dependent on vertex density and ensures uniform scattering also when $q$ is low.

\subsection{\CGnew{Parameter selection}}
\label{ssec:param}

\CGnew{As we showed in Section~\ref{sec:gaussian_dictionary}, \textbf{\PCGAU{}} computation requires two parameters, namely the number $q$ and the amplitude $\sigma$ of the Gaussian functions. In this Section, we analyze the impact of such parameters on the accuracy of the final point-wise map, and we show that we can exploit the metrics presented in Section~\ref{ssec:discrimination_localization} and~\ref{ssec:locality_preservation} to select appropriate values for $\sigma$ and $q$, without any knowledge of the ground-truth correspondence.}

\subsubsection{Amplitude of Gaussian functions $\sigma$}
\begin{figure}[t]
    \centering
    \subfloat[error (GT $C$)]{\includegraphics[width=0.32\linewidth]{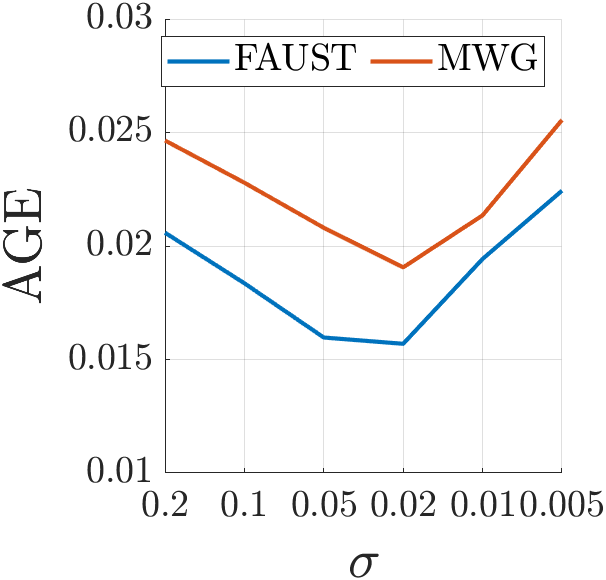} \label{sfig:selection_sigma:error_gt}} \hfill
    \subfloat[error (NO17 $C$)]{\includegraphics[width=0.32\linewidth]{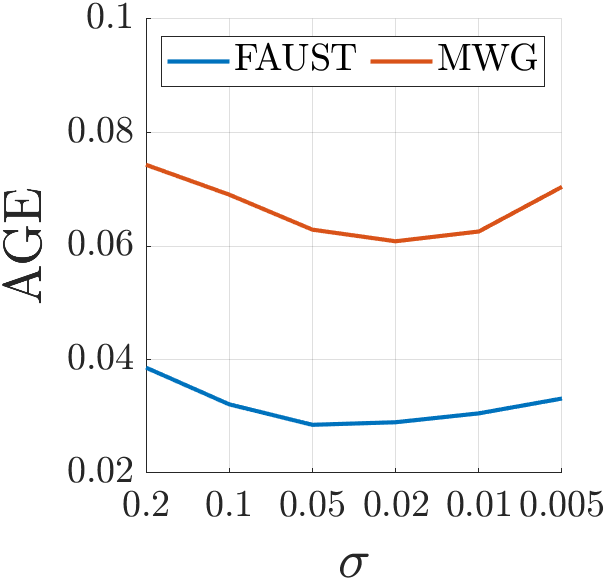} \label{sfig:selection_sigma:error_no17}} \hfill
    \subfloat[locality preservation]{\includegraphics[width=0.32\linewidth]{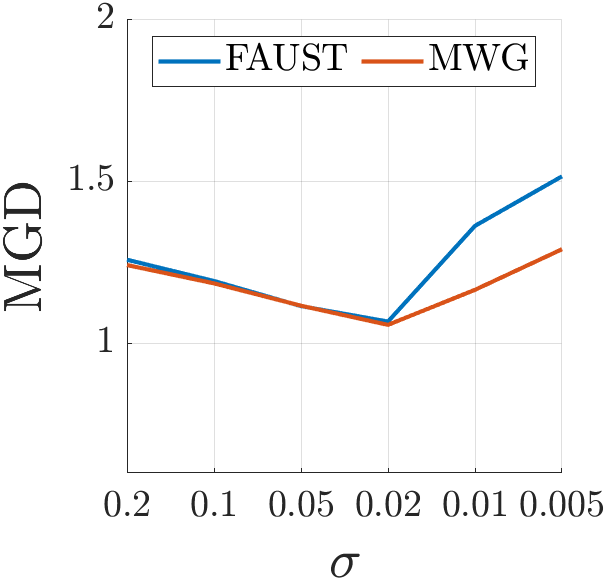} \label{sfig:selection_sigma:loc_pres}} \hfill
    \caption[Comparison of point-wise error and locality preservation among different $\sigma$]{\CGnew{Comparison of point-wise error \protect\subref{sfig:selection_sigma:error_gt} (ground-truth), \protect\subref{sfig:selection_sigma:error_no17} (NO17), and locality preservation \protect\subref{sfig:selection_sigma:loc_pres} for different values of $\sigma$. MGD is an indicator of the error at test time, thus, we can exploit it to select the best value of $\sigma$.}}
    \label{fig:selection_sigma}
\end{figure}
\begin{figure}[t]
    \centering
    \subfloat[error (GT $C$)]{\includegraphics[width=0.32\linewidth]{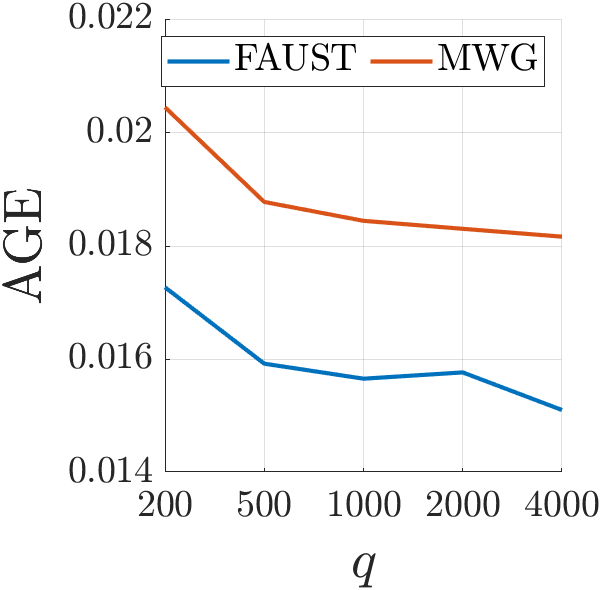} \label{sfig:selection_q:error_gt}} \hfill
    \subfloat[error (NO17 $C$)]{\includegraphics[width=0.32\linewidth]{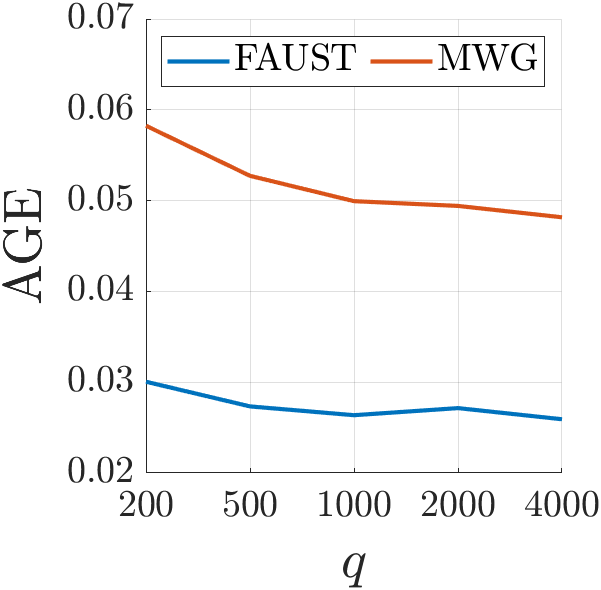} \label{sfig:selection_q:error_no17}} \hfill
    \subfloat[locality preservation]{\includegraphics[width=0.32\linewidth]{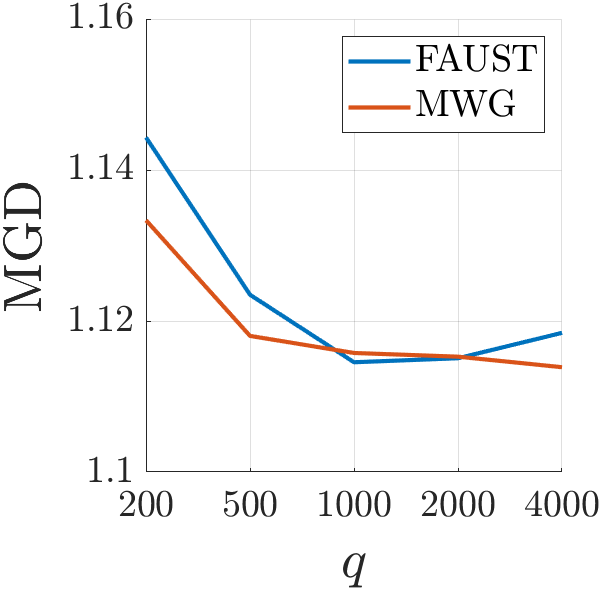} \label{sfig:selection_q:loc_pres}} \hfill
    \caption[Comparison of point-wise error and locality preservation among different $q$]{\CGnew{Comparison of point-wise error \protect\subref{sfig:selection_sigma:error_gt} (ground-truth), \protect\subref{sfig:selection_sigma:error_no17} (NO17), and locality preservation \protect\subref{sfig:selection_sigma:loc_pres} among different values of $q$. Average on 30 pairs from FAUST and MWG datasets.}}
    \label{fig:selection_q}
\end{figure}
\CGnew{As shown in Section~\ref{sec:gaussian_dictionary}, the parameter $\sigma$ sets the amplitude of the Gaussian functions: the lower the value of $\sigma$ and the more the Gaussian function is localized around its center point.} 

\CGnew{Figure~\ref{sfig:selection_sigma:error_gt} and~\ref{sfig:selection_sigma:error_no17} present the average geodesic error of point-wise maps ($y$-axis) obtained using different values of $\sigma$ ($x$-axis). For each value of $\sigma$ \textbf{\PCGAU{}} is computed and then tested both estimating $C$ using the ground-truth correspondence, as done in Section~\ref{ssec:exp_gt}, and using product preservation method \cite{nogneng17}, as done in Section~\ref{ssec:exp_no17}. The plots show that values of $\sigma$ that are too large and too small result in a sensible decrease in accuracy in the maps obtained. Therefore it is important to choose a value of $\sigma$ near the mean. Figure~\ref{sfig:selection_sigma:loc_pres} shows the value of MGD obtained by our basis computed with different values of $\sigma$. We can see that the value of $\sigma$ that minimizes $\mgd$ is also optimal or nearly optimal for the point-wise error. Since we compute $\mgd$ for every single mesh and, thus, without any knowledge about the ground truth correspondence, we can adopt this analysis in real scenarios where the correspondence between different meshes is unknown. Here, we only consider $\mgd$, but we obtain similar results with $\egdc$ and $\discr$, which behave similarly to $\mgd$.}

\subsubsection{\CGnew{Number of Gaussian functions $q$}}
\CGnew{The parameter $q$ determines the number of points selected on the mesh surface and, thus, the number of Gaussian functions.
Similarly to the previous paragraph, Figure~\ref{fig:selection_q} shows both the point-wise error and the value of $\mgd$ for \textbf{\PCGAU{}} ($y$-axis) computed using different values of $q$ ($x$-axis). We can still observe a strong relationship between the metric and the error obtained, thus allowing us to use $\mgd$ also for this parameter. We remark that the error decreases as $q$ increases, with less noticeable effects after $q=1000$. However, the time needed to compute our basis increases more than linearly with $q$. Therefore, $q=1000$, i.e. the value used in our experiments, represents a good trade-off between accuracy and efficiency.}

\CGnew{Note, at last, that the predictive power of the metrics defined in Section~\ref{ssec:discrimination_localization} and~\ref{ssec:locality_preservation} on the point-wise error is an additional confirmation of their relevance. This relationship means that they assess the properties of the embedding space relevant to the conversion to point-wise maps.
}

\section{Conclusions}
\label{sec:conclusions}
\CGnew{We presented the procedure to build a new basis for the space of real-valued functions defined on a mesh. Our framework, namely \ourname{}, starts by selecting a dictionary of informative functions for the mesh and then extracts an orthonormal basis by applying the principal component analysis. The output of \ourname{} can replace LB at no cost in virtually any functional map pipeline to target the shape matching task. In particular, we focused on \PCGAU{}, the basis obtained by selecting a dictionary of Gaussians, which are easy to compute and give rise to desirable properties.} Compared to the eigenfunctions of the Laplace-Beltrami operator, which are the standard basis for functional maps, the energy of \PCGAU is distributed more uniformly on the mesh. The resulting embedding space for the vertices is overall more amenable to point-wise conversion.
\CGnew{We extended our analyses to different compositions of the initial dictionary, evaluating their performances on several established datasets and using different methods for estimating the functional map. We showed that substituting LB with \ourname{}, for some specific choices of the initial dictionary, leads to far superior results in the accuracy of the obtained point-wise map. We also analyzed the dependency of the matching accuracy on the value of the parameters required by \PCGAU{}, showing a simple method to choose them in real-case scenarios.}

Although the results are pretty impressive, we experienced that \ourname{} has some limitations: \emph{i}) it suffers when there are significant errors in the estimation of the functional map; \emph{ii}) the entire framework has been defined for meshes, while we did not explore its applicability to point clouds, neither considering meshes extracted from dense point clouds nor directly working on 3D points without connectivity; \emph{iii}) we believe that even if the embedding from \PCGAU{} is uniform, this is not optimal and could be improved by explicitly enforcing this property.

A first interesting direction to explore in the future is to test the ability of our basis to represent different signals defined on the surfaces as done in~\cite{melzi2019zoomout}, to evaluate if the functional representation can also benefit from the more uniform distribution of the energy of our basis. \CGnew{A second direction we only partially addressed is to evaluate other possible dictionary compositions as inputs of our procedure. We saw that some alternatives could provide better results compared to \PCGAU{}, even if only on some test sets.} Finally, we would investigate how it is possible to integrate \ourname{} in data-driven procedures inspired by the functional maps framework, such as~\cite{litany17deep,donati2020deep,marin2020correspondence}.

\CGnew{\section*{Acknowledgements}
This paper is supported by the PNRR-PE-AI FAIR project funded by the NextGeneration EU program. This work has been partially supported by the SAPIENZA BE-FOR-ERC 2020 Grant (NONLINFMAPS). 
We gratefully acknowledge the support of NVIDIA Corporation with the
RTX A5000 GPUs granted through the Academic Hardware Grant Program to the Politecnico di Milano for the project "Direct Experience in Deep Learning" and to the University of Milano-Bicocca for the project "Learned representations for implicit binary operations on real-world 2D-3D data".
}
\bibliographystyle{abbrv}
\bibliography{fmapbib}
\end{document}